\documentclass[nologo,entropy,article,submit,pdftex,moreauthors]{Definitions/mdpi} 
\firstpage{1} 
\pubvolume{1}
\issuenum{1}
\articlenumber{0}
\pubyear{2026}
\copyrightyear{2026}
\datereceived{ } 
\daterevised{ } 
\dateaccepted{ } 
\datepublished{ } 

\Title{Fluctuating Hydrodynamics of Quantum Fluids from the Nonlinear Fokker-Planck Equation}

\Author{Mingjian Gu $^{1}$\orcidA{}, Zhaoyu fei $^{1,}$* and Xiaoguang Wang $^{1}$}

\AuthorNames{Firstname Lastname, Firstname Lastname and Firstname Lastname}

\address{%
$^{1}$ \quad Zhejiang Key Laboratory of Quantum State Control and Optical Field Manipulation, Department of Physics, Zhejiang Sci-Tech University, Hangzhou, Zhejiang 310018, China; feizhaoyu@zstu.edu.cn\\}

\corres{Correspondence: feizhaoyu@zstu.edu.cn}

	\abstract{Starting from the nonlinear Fokker-Planck equation (NFPE), we establish the fluctuating hydrodynamics for quantum fluids. We also investigate the corresponding stochastic thermodynamics and derive the associated fluctuation theorems. We then apply this framework to both fermionic and bosonic systems. For the ideal Fermi gas, we study the electrical conduction and reveal a resistance proportional to the product of temperature and the density of states at the Fermi surface. For bosonic fluids, we analyze the relaxation time characterizing the critical slowing down phenomenon and identify three distinct scaling behaviors corresponding to the non-interacting, repulsive interaction, and attractive interaction regimes. Overall, this work provides a experimentally verifiable phenomenological description for the dynamics of quantum fluids and demonstrates the applicability of the nonlinear Fokker-Planck equation in characterizing such systems.}


\begin{document}

\setcounter{section}{-1} 


\section{Introduction}

The traditional linear Fokker-Planck equation has long served as a cornerstone in non-equilibrium statistical mechanics, achieving remarkable success in describing classical Brownian motion and macroscopic transport phenomena \cite{risken1989fokker, zwanzig2001nonequilibrium}. However, this standard linear framework encounters fundamental limitations when extended to quantum systems, as it lacks the inherent capability to incorporate quantum statistics and particle interactions, both of which inevitably break the linearity of the evolution equations~\cite{frank2005nonlinear, chavanis2008nonlinear}. To overcome these fundamental challenges, the NFPE has been extensively developed as a powerful framework~\cite{kadanoff2000statistical,kaniadakis2001h,frank2005nonlinear,chavanis2008nonlinear,falasco2025macroscopic}. Unlike linear equations that relax to the Boltzmann-Gibbs distribution, these nonlinear equations admit non-Boltzmann distributions as their steady states. Consequently, the NFPE framework has found extensive applications in modeling complex systems, such as investigating the collective dynamics of active Brownian particle~\cite{menzel2016dynamical,wittmann2016active} and colloidal fluids~\cite{wensink2008aggregation,wittkowski2011dynamical} and biological chemotaxis~\cite{chavanis2010stochastic, chavanis2014random} for classical systems, as well as characterizing equilibration processes in fermionic and bosonic systems~\cite{wolschin1982equilibration, wolschin2018equilibration} and describing electron collisions in dense plasmas~\cite{daligault2016quantum} for quantum systems.

Motivated by its inherent capability to incorporate both quantum statistics and many-body interactions, we formulate a macroscopic fluctuating hydrodynamics for quantum fluids based on the NFPE. In the classical regime, fluctuating hydrodynamics has been systematically developed through the Dean-Kawasaki framework. Specifically, Kawasaki combined the local equilibrium approximation with density functional theory to describe coarse-grained density fields in interacting fluids~\cite{kawasaki1994stochastic,kawasaki1998microscopic}, while Dean derived the corresponding Langevin equation for microscopic density fields with multiplicative noise, capturing long-wavelength scaling behaviors in classical non-equilibrium fluids~\cite{dean1996langevin,te2020classical,illien2025dean}. Currently, existing quantum applications of the NFPE primarily focus on density evolution and spatial diffusion. In this paper, starting from the phase-space formalism of NFPE~\cite{fei2023quantum}, we systematically extend this framework to a macroscopic fluctuating hydrodynamic description. Specifically, by invoking the local equilibrium assumption, we derive a set of phenomenological fluctuating hydrodynamic equations. Compared to other quantum fluid descriptions—such as the quantum Boltzmann equation, Gross-Pitaevskii methods~\cite{zaremba1999dynamics,bijlsma2000condensate}, and discrete exclusion models~\cite{bertini2015macroscopic}—which typically demand highly complex mathematical treatments to extract macroscopic equations, our NFPE-based formulation circumvents these intricate derivations. Instead, it provides a direct framework to simultaneously capture macroscopic momentum transport and quantum statistical fluctuations.

Building on these hydrodynamic equations, we further construct a consistent stochastic thermodynamic framework at the macroscopic field level. By defining trajectory-level work, heat, and entropy production along the fluctuating hydrodynamic fields, we formulate the first and second laws of thermodynamics, as well as the integral and detailed fluctuation theorems far from equilibrium~\cite{sekimoto2010stochastic,jarzynski2012equalities,crooks1998nonequilibrium,seifert2012stochastic}. Specifically, unlike abstract phase-space distributions that are difficult to measure directly, our macroscopic framework governs the spatial evolution of physical fields that are potentially accessible to experimental verification, for instance, via high-resolution optical imaging~\cite{bakr2009quantum}.

To demonstrate the physical validity of our theoretical framework, we apply the derived hydrodynamic equations to investigate two representative systems across distinct quantum statistical regimes. First, we examine the electrical conduction of an ideal Fermi gas driven by a external electric field. While our macroscopic momentum balance equation naturally reduces to the Drude model, it offers a fundamental advantage: unlike the conventional Drude approach where temperature dependence of resistance must be phenomenologically introduced, our framework directly derives a resistance proportional to the product of temperature and the density of states at the Fermi surface. This result is consistent with electron-phonon scattering theories~\cite{mott1936electrical} and experimental observations~\cite{ashcroft1976solid}. Second, we explore the relaxation dynamics of overdamped bosonic fluids approaching the Bose-Einstein condensation (BEC) threshold from above the critical temperature. By deriving a nonlinear diffusion equation and performing a linear stability analysis for the local chemical potential, we quantitatively characterize the critical slowing down of the macroscopic relaxation near the phase transition point~\cite{hohenberg1977theory}. Specifically, we demonstrate that the scaling behaviors of the divergence of the macroscopic relaxation time are fundamentally governed by two-body interactions, exhibiting distinct behaviors across non-interacting, repulsive interaction, and attractive interaction  regimes.

The rest of this paper is organized as follows. In Section~\ref{sec:hydrodynamics}, we pass from the phase-space description to the hydrodynamic description by invoking the local equilibrium assumption to calculate the velocity moments of the NFPE. In Section~\ref{sec:functional_fpe}, we introduce macroscopic fluctuations and formulate the hydrodynamic functional Fokker--Planck equation. Building on this, Section~\ref{sec:thermodynamics} develops the corresponding stochastic thermodynamics and macroscopic fluctuation theorems. In Section~\ref{sec:applications}, we apply our theory to two paradigmatic quantum systems: we first investigate the electric current in a circuit modeled as an ideal Fermi gas in Section~\ref{subsec:circuit_current}, and then analyze the stability and critical slowing down near the critical temperature of BEC in Section~\ref{subsec:bec_criticality}. Finally, we present our conclusions and outlook in Section~\ref{sec:conclusion}.
\section{From Phase-Space Description to Hydrodynamics Description} \label{sec:hydrodynamics}

Let us consider a quantum fluid composed of $N$ identical particles. By employing the semiclassical approximation and introducing the single-particle phase-space distribution function $f(\mathbf{r}, \mathbf{p}, t)$, where $\mathbf{r} = (r_1, \dots, r_d)$ and $\mathbf{p} = (p_1, \dots, p_d)$ denote the $d$-dimensional spatial and microscopic momentum vectors respectively, the macroscopic total energy functional of the system is formulated as \cite{fei2024temperature, huang2008statistical}
\begin{equation}\label{eq:total_energy}
	E[f] = \int \frac{\mathrm{d}\mathbf{r} \mathrm{d}\mathbf{p}}{h^d} \, f(\mathbf{r}, \mathbf{p}, t) \left[\ \frac{\mathbf{p}^2}{2m} + m \Phi(\mathbf{r},t) \right].
\end{equation}
The energy functional is governed by the phase-space coordinates, consisting of a kinetic part parameterized by the microscopic momentum $\mathbf{p}$, and a potential part $m\Phi$. Specifically, the potential $\Phi$ can be systematically decomposed into an external potential $\Phi_{\mathrm{ext}}(\mathbf{r}, t)$ and a mean-field effective potential $\Phi_{\mathrm{int}}(\mathbf{r}, t)$, i.e., $\Phi(\mathbf{r}, t) = \Phi_{\mathrm{ext}}(\mathbf{r}, t) + \Phi_{\mathrm{int}}(\mathbf{r}, t)$, where the latter is expressed through the phase-space distribution \cite{chavanis2008nonlinear}
\begin{equation}\label{eq:effective_potential}
	\Phi_{\mathrm{int}}(\mathbf{r}, t) = \int \frac{\mathrm{d}\mathbf{r}' \mathrm{d}\mathbf{p}'}{h^d} \, f(\mathbf{r}', \mathbf{p}', t) v(\mathbf{r} - \mathbf{r}').
\end{equation}
Here, $m$ is the mass of a single particle, $h$ denotes the Planck constant, and $v(\mathbf{r} - \mathbf{r}')$ represents the two-body interaction potential between particles.

To account for the non-equilibrium dynamics driven by both stochastic thermal fluctuations and quantum statistics, the time evolution of the phase-space density is governed by a non-linear stochastic Fokker-Planck equation \cite{fei2023quantum}, explicitly given by
\begin{equation} \label{eq:sFPE}
	\frac{\partial f}{\partial t} = - \frac{\mathbf{p}}{m} \cdot \nabla_{\mathbf{r}} f + m \nabla_{\mathbf{r}} \Phi \cdot \nabla_{\mathbf{p}} f + \nabla_{\mathbf{p}} \cdot \mathbf{j}_t,
\end{equation}
where, the first two terms on the RHS represent the reversible Liouville evolution of the system in the absence of environmental coupling. The last term captures the irreversible dynamics arising from the coupling to a thermal reservoir at temperature $T_r$. This interaction induces a phase-space flux density $\mathbf{j}_t$, which consists of a deterministic dissipative current and a stochastic noise term
\begin{equation}
	\mathbf{j}_t = m \gamma f(1 + \epsilon f) \nabla_{\mathbf{p}} \left( \frac{\delta F[f]}{\delta f} \right) + \boldsymbol{\eta}.
\end{equation}
This structure embodies a generalized Onsager relation, where the irreversible flux $\mathbf{j}_t$ equals the product of an Onsager coefficient $m\gamma f(1+\epsilon f)$ and a thermodynamic driving force $\nabla_{\mathbf{p}}(\delta F/\delta f)$, where $\gamma$ denotes the damping coefficient and $\delta$ represents the functional variation. Crucially, the Onsager coefficient incorporates the quantum-statistical factor $(1+\epsilon f)$, where $\epsilon = 1, -1, 0$ represents bosons, fermions, and distinguishable particles, respectively. This factor explicitly captures the macroscopic manifestation of quantum statistics: it yields enhanced statistical effects for bosons and suppressed for fermions \cite{kaniadakis1994classical}.
The thermodynamic driving force is derived from the macroscopic nonequilibrium free energy functional $F[f] = E[f] - T_r S[f]$, where $T_r$ is the temperature of the reservoir. Here, $S[f]$ is the Boltzmann entropy for a system of identical particles, given by
\begin{equation}
	S[f] = k_B \int \frac{\mathrm{d}\mathbf{r} \mathrm{d}\mathbf{p}}{h^d} \, \left[ -f \ln f + \epsilon^{-1}(1 + \epsilon f) \ln(1 + \epsilon f) \right].
\end{equation}
The stochastic dynamics are driven by a white noise $\boldsymbol{\eta}$ in phase space, which originates from the random collisions between the system and the environment, satisfying $\langle \boldsymbol{\eta} \rangle = \mathbf{0},	\langle \eta_{i}(\mathbf{r}, \mathbf{p}, t) \eta_{j}(\mathbf{r}', \mathbf{p}', t') \rangle = 2m\gamma k_B T_r h^d f (1 + \epsilon f) \delta_{ij} \delta(\mathbf{r} - \mathbf{r}') \delta(\mathbf{p} - \mathbf{p}') \delta(t - t')$ \cite{dean1996langevin, du2025repulsive, du2024correlation}. The noise intensity is fundamentally constrained by the generalized fluctuation-dissipation theorem to ensure that the system relaxes to a stable quasi-equilibrium state. 
Consequently, the non-linearity of the dynamics stems from both the quantum-statistical factor in the Onsager coefficient and the entropic contribution to the thermodynamic force through the free energy functional.

We partition the system into fluid cells that are macroscopically infinitesimal yet microscopically large enough to contain a statistically meaningful number of particles. To extract the macroscopic transport implications of Eq.~\eqref{eq:sFPE}, we exploit the timescale separation between fast momentum relaxation and slow spatial evolution. 
Under the local equilibrium assumption, the non-equilibrium phase-space density is characterized by spatiotemporally varying thermodynamic variables, and is thus approximated by the local equilibrium form \cite{glansdorff1973thermodynamic, kreuzer1981nonequilibrium, de2013non}
\begin{equation}\label{eq:evolution_reduced}
	f(\mathbf{r},\mathbf{p},t) = \frac{1}{\exp\left\{ \beta(\mathbf{r},t) \left[ \frac{\left(\mathbf{p} - m\mathbf{u}(\mathbf{r},t)\right)^2}{2m} + m\Phi(\mathbf{r},t) - \mu(\mathbf{r},t) \right] \right\} - \epsilon},
\end{equation}
where $\beta(\mathbf{r},t) = [k_B T(\mathbf{r},t)]^{-1}$ is the local inverse temperature, $\mathbf{u}(\mathbf{r},t)$ denotes macroscopic flow velocity and $\mu(\mathbf{r},t)$ represents chemical potential. These local thermodynamic parameters are determined by the balance equations, which will be explicitly formulated later.

The stochastic Fokker-Planck equation~\eqref{eq:sFPE} combined with the local equilibrium distribution~\eqref{eq:evolution_reduced} leads to an effective kinetic equation
\begin{equation}\label{eq:Kramer_equation}
	\frac{\partial f}{\partial t} + \frac{\mathbf{p}}{m} \cdot \nabla_{\mathbf{r}} f - m \nabla_{\mathbf{r}} \Phi \cdot \nabla_{\mathbf{p}} f 
	= \nabla_{\mathbf{p}} \cdot \mathbf{j}_t^*,
\end{equation}
where the flux after imposing the local equilibrium assumption takes the form
\begin{equation}
	\mathbf{j}_t^* =\gamma f(1 + \epsilon f) \left[ \mathbf{p} - \frac{T_r}{T} (\mathbf{p} - m\mathbf{u}) \right] + \boldsymbol{\eta}.
\end{equation}

Macroscopic hydrodynamic densities are obtained from the phase-space density $f$ via integration over the momentum space $\mathrm{d}\mathbf{p} / h^d$.
Specifically, the mass density $\rho$, the macroscopic flow velocity $\mathbf{u}$, the specific internal energy $u$, specific total energy $e$, and specific entropy $s$ are given by
\begin{align}
	\rho &\equiv \int \frac{\mathrm{d}\mathbf{p}}{h^d} \, f ,\\
	\rho \mathbf{u} &\equiv \int \frac{\mathrm{d}\mathbf{p}}{h^d} \, f \frac{\mathbf{p}}{m} ,\\
	\rho u &\equiv \int \frac{\mathrm{d}\mathbf{p}}{h^d} \, f \frac{(\mathbf{p}-m\mathbf{u})^2}{2m} , \label{eq:internal_energy_def}\\
	\rho e &\equiv \int \frac{\mathrm{d}\mathbf{p}}{h^d} \, f \left( \frac{\mathbf{p}^2}{2m} + m\Phi \right) , \label{eq:total_energy_def}\\
	\rho s &\equiv -k_B \int \frac{\mathrm{d}\mathbf{p}}{h^d} \left[ f \ln f - \epsilon^{-1}(1 + \epsilon f) \ln(1 + \epsilon f) \right] . \label{eq:entropy_def}
\end{align}

Taking the zeroth, first, and second momentum moments of Eq.~\eqref{eq:Kramer_equation} yields the macroscopic hydrodynamic equations governing the exact local density, momentum, and total energy of the fluid, respectively
\begin{align}
	\frac{\partial \rho}{\partial t} + \nabla \cdot (\rho \mathbf{u}) &= 0, \label{eq:continuity} \\
	\rho \left[ \frac{\partial \mathbf{u}}{\partial t} + (\mathbf{u} \cdot \nabla)\mathbf{u} \right] &= -\nabla p - \rho \nabla \Phi - \xi \rho \mathbf{u} + \boldsymbol{\chi}, \label{eq:momentum} \\
	\rho \left[ \frac{\partial e}{\partial t} + (\mathbf{u} \cdot \nabla)e \right] &= -\nabla \cdot (p\mathbf{u}) + \rho \frac{\partial \Phi}{\partial t} - \xi \rho \mathbf{u}^2 - \gamma k_B \frac{\rho}{m} (T - T_r) + \mathbf{u} \cdot \boldsymbol{\chi} + \Gamma, \label{eq:total_energy}
\end{align}
where $p = \frac{1}{d} \int \frac{\mathrm{d}\mathbf{p}}{h^d} f \frac{(\mathbf{p}-m\mathbf{u})^2}{m}$ is the pressure, arising from the relative motion of particles with respect to the macroscopic flow. Here, the effective friction coefficient is defined as 
\begin{equation} \label{eq:xi}
	\xi \equiv \frac{\gamma k_B T}{m c_T^2},
\end{equation}
where $c_T^2 \equiv (\partial p / \partial \rho)_T$ denotes part of the isothermal sound speed\footnote{It is important to note that $c_T^2$ here represents solely the relative motion pressure to the sound speed. The standard total sound speed would require incorporating the additional pressure contributions arising from macroscopic interparticle interactions.}. The macroscopic stochastic force $\boldsymbol{\chi} \equiv - \int \frac{\mathrm{d}\mathbf{p}}{h^d} \boldsymbol{\eta}$ and the stochastic heat exchange $\Gamma \equiv - \int \frac{\mathrm{d}\mathbf{p}}{h^d} \frac{(\mathbf{p} - m\mathbf{u})}{m} \cdot \boldsymbol{\eta}$ are two independent noises. They satisfy $\langle \boldsymbol{\chi} \rangle = \mathbf{0}$ and $\langle \Gamma \rangle = 0$, with their variances given by $\langle \chi_i(\mathbf{r}, t) \chi_j(\mathbf{r}', t') \rangle = 2 \xi k_B T_r \rho \, \delta_{ij} \delta(\mathbf{r} - \mathbf{r}') \delta(t - t')$ and $\langle \Gamma(\mathbf{r}, t) \Gamma(\mathbf{r}', t') \rangle = 2 \gamma k_B^2 T_r T \frac{\rho}{m} \delta(\mathbf{r} - \mathbf{r}') \delta(t - t')$.

While the first three moment equations collectively determine the macroscopic hydrodynamic dynamics of the system, they do not explicitly reveal its underlying thermodynamic structure. To elucidate the non-equilibrium thermodynamic properties, we derive the balance equations for the specific internal energy and the specific entropy.
By evaluating the corresponding moment associated with the definition in Eq.~\eqref{eq:internal_energy_def}, we obtain the macroscopic balance equation for the specific internal energy $u$
\begin{equation} \label{eq:internal_energy}
	\rho \left[ \frac{\partial u}{\partial t} + (\mathbf{u} \cdot \nabla)u \right] = -p \nabla \cdot \mathbf{u} - \gamma k_B \frac{\rho}{m} (T - T_r) + \Gamma.
\end{equation}
The second term on the RHS of Eq.~\eqref{eq:internal_energy} represents Newton's law of cooling, while the variance of the stochastic term $\Gamma$ is proportional to $T T_r$, consistent with the diffusion coefficient for temperature fluctuations~\cite{fei2024temperature}.
The time evolution of entropy follows directly from its definition~\eqref{eq:entropy_def} together with the kinetic equation~\eqref{eq:Kramer_equation}
\begin{equation} \label{eq:entropy}
	\rho \left[ \frac{\partial s}{\partial t} + (\mathbf{u} \cdot \nabla)s \right] = -\gamma k_B \frac{\rho}{m} \frac{T - T_r}{T} + \frac{\Gamma}{T}.
\end{equation}
Equation~\eqref{eq:entropy} exhibits an endoreversible structure: the entropy change of the fluids equals the heat flux divided by the system temperature $T$, with the heat exchange is driven by the temperature difference $T-T_r$ \cite{curzon1975efficiency}. This is a direct consequence of the local equilibrium assumption, which forces all internal processes to be reversible.

We note that the hydrodynamic equations derived above contain no viscous stress tensor and no internal heat flux. This follows inevitably from neglecting gradient corrections in the local equilibrium assumption. This assumption thus restricts the present model to a friction-dominated regime where momentum and energy relaxation are exclusively controlled by the external reservoir rather than by internal viscous dissipation. For a framework that explicitly incorporates the viscous stress tensor and internal heat flux, we refer the reader to Ref.~\cite{forastiere2024dissipation}.

By the way, the distinction between It\^{o} and Stratonovich integration schemes in the stochastic process becomes immaterial in the hydrodynamic limit. Macroscopically, the relative amplitude of the thermal fluctuations scales inversely with the square root of the particle number within a fluid cell, i.e., $\sim \mathcal{O}(N^{-1/2})$ \cite{landau1970course, donev2014reversible, chavanis2019generalized}. According to the system size expansion, the integration scheme ambiguity manifests as a noise-induced spurious drift strictly of order $\mathcal{O}(N^{-1})$ \cite{van1983stochastic, chavanis2019generalized}. Since our macroscopic framework retains terms up to $\mathcal{O}(N^{-1/2})$ to capture the leading-order thermal fluctuations, the $\mathcal{O}(N^{-1})$ spurious drift is strictly a higher-order small quantity relative to the retained noise. Consequently, this term vanishes in the thermodynamic limit, which justifies dropping the integration scheme ambiguity in the derivations of the macroscopic transport equations.
\section{Hydrodynamic Functional Fokker-Planck Equation} \label{sec:functional_fpe}

The hydrodynamic equations derived in the previous section describe the macroscopic dynamics as a set of stochastic field equations. 
While this Langevin-type description directly tracks the noisy trajectories of the fields, it is also powerful to adopt an equivalent probabilistic viewpoint, i.e., a functional Fokker-Planck equation that governs the probability distribution of the macroscopic field configurations.
To construct this description, we must first identify the independent macroscopic variables. 
Under the local equilibrium assumption, the system at each space-time point is fully characterized by the mass density $\rho(\mathbf{r},t)$, the flow velocity $\mathbf{u}(\mathbf{r},t)$, and the specific entropy density $s(\mathbf{r},t)$. 
All other thermodynamic quantities, such as the internal energy $u$, temperature $T$, and pressure $p$, follow from $\rho$ and $s$ via the equation of state. 

We take the independent macroscopic fields as $\boldsymbol{\Psi} = (\rho, \mathbf{u}, s)$, the probability distribution functional $P[\boldsymbol{\Psi}; t]$ over the hydrodynamic fields satisfies the functional Fokker-Planck equation (see Appendix~\ref{app:FFP} for the detailed derivation)
\begin{equation} \label{eq:functional_FPE}
	\frac{\partial P}{\partial t} = \int \mathrm{d}\mathbf{r} \, \{F, P\} + \int \mathrm{d}\mathbf{r} \left( \frac{\delta}{\delta \mathbf{u}} \cdot \frac{\xi P}{\rho} \frac{\delta}{\delta \mathbf{u}} + \frac{\delta}{\delta s} \frac{\gamma k_B P}{\rho m T} \frac{\delta}{\delta s} \right) \left( F + k_B T_r \ln P \right).
\end{equation}
The RHS of Eq.~\eqref{eq:functional_FPE} contains two distinct contributions. The first term, involving the Poisson bracket with the free energy functional $F$, describes the conservative and reversible part of the hydrodynamic evolution. The second term, featuring the functional dissipation operators, captures the irreversible processes that drive the system toward thermodynamic equilibrium.

To specify the structure of the conservative dynamics in the first term, we construct the non-equilibrium free energy functional $F = E - T_r S$. It is obtained by integrating the corresponding local total energy and entropy densities over the spatial domain
\begin{align}
	E &= \int \mathrm{d}\mathbf{r} \, \rho e = \int \mathrm{d}\mathbf{r} \, \rho \left( \frac{1}{2}\mathbf{u}^2 + \Phi + u \right), \label{eq:total_E} \\
	S &= \int \mathrm{d}\mathbf{r} \, \rho s. \label{eq:total_S}
\end{align}
By utilizing the local thermodynamic identity $\mathrm{d}u = T\mathrm{d}s + (p/\rho^2)\mathrm{d}\rho$, the functional derivatives of the free energy $F$ with respect to the fundamental hydrodynamic fields $\boldsymbol{\Psi}$ can be explicitly evaluated as
\begin{equation} \label{eq:F_derivatives}
	\frac{\delta F}{\delta \rho} = \frac{1}{2}\mathbf{u}^2 + \Phi + u + \frac{p}{\rho} - T_r s, \quad
	\frac{\delta F}{\delta \mathbf{u}} = \rho \mathbf{u}, \quad
	\frac{\delta F}{\delta s} = \rho (T - T_r).
\end{equation}
These functional derivatives act as the conjugate thermodynamic driving forces governing the macroscopic evolution.
The operation $\{A, B\}$ denotes the non-canonical Poisson bracket \cite{edwards1991non,muschik2001sketch,grmela1997dynamics,ottinger1997dynamics}, explicitly given by
\begin{equation} \label{eq:poisson_bracket}
	\{A, B\} = \frac{\delta A}{\delta \mathbf{u}} \cdot \left( \frac{\nabla s}{\rho} \frac{\delta}{\delta s} - \nabla \frac{\delta}{\delta \rho} \right) B 
	- \frac{\delta B}{\delta \mathbf{u}} \cdot \left( \frac{\nabla s}{\rho} \frac{\delta}{\delta s} - \nabla \frac{\delta}{\delta \rho} \right) A.
\end{equation}
We note that the bracket is driven directly by the functional $F$ rather than the total energy $E$. 

In the presence of a confining potential $\Phi$, the system is spatially bounded. Consequently, the system relaxes toward a stationary state. The functional Fokker-Planck equation then admits a stationary equilibrium solution
\begin{equation} \label{eq:equilibrium_P}
	P_{s}[\boldsymbol{\Psi}] = \frac{1}{\mathcal{Z}} e^{-\beta_r F[\boldsymbol{\Psi}]} \, \delta(N - N[\rho]),
\end{equation}
where $\mathcal{Z} = \int \mathcal{D}\boldsymbol{\Psi} e^{-\beta_r F[\boldsymbol{\Psi}]} \delta(N - N[\rho])$ is the partition functional and $N[\rho] = \int \mathrm{d}\mathbf{r} \, \rho$. The Dirac delta enforces the strict conservation of the total particle number. Consistent with the theory of equilibrium fluctuations \cite{Mishin2015,fei2023quantum}, $P_{s}$ characterizes a canonical quasi-equilibrium state.

To explore a broader physical scenario including a non-conservative force $\mathbf{F}_{\mathrm{nc}}$, we now consider a distinct setting where the external confining potential is absent ($\Phi_{\mathrm{ext}} = 0$) and the system is subjected to periodic boundary conditions. In this configuration, the functional Fokker-Planck equation modifies to include an additional drift term (see Appendix~\ref{app:FFP} for the detailed derivation)
\begin{equation} \label{eq:driven_FPE}
	\frac{\partial P}{\partial t} = \int \mathrm{d}\mathbf{r} \, \{F, P\} - \int \mathrm{d}\mathbf{r} \, \mathbf{F}_{\mathrm{nc}} \cdot \frac{\delta P}{\delta \mathbf{u}} + \int \mathrm{d}\mathbf{r} \left( \frac{\delta}{\delta \mathbf{u}} \cdot \frac{\xi P}{\rho} \frac{\delta}{\delta \mathbf{u}} + \frac{\delta}{\delta s} \frac{\gamma k_B P}{\rho m T} \frac{\delta}{\delta s} \right) \left( F + k_B T_r \ln P \right).
\end{equation}
When the non-conservative force is constant, it injects mechanical power into the fluid, preventing the system from relaxing to a static equilibrium and driving the system instead into a non-equilibrium steady state sustained by a macroscopic flow $\langle \mathbf{u} \rangle$

\begin{equation} \label{eq:steady_P}
	P_s[\boldsymbol{\Psi}] = \frac{1}{\widetilde{\mathcal{Z}}} 
	\exp\left[ \frac{S[\boldsymbol{\Psi}]}{k_B} - \beta_r \int \mathrm{d}\mathbf{r} \, \rho \left( \frac{1}{2} \big(\mathbf{u} - \langle \mathbf{u} \rangle\big)^2 + \Phi_{\mathrm{int}} + u \right) \right]
	\delta(N - N[\rho]),
\end{equation}
where $\widetilde{\mathcal{Z}}$ is the normalization factor for the steady state. The macroscopic drift velocity is determined by the uniform non-conservative driving force and the effective friction, $\langle \mathbf{u} \rangle = \mathbf{F}_{\mathrm{nc}} / \xi$.

\section{Stochastic Thermodynamics} \label{sec:thermodynamics}

To establish the thermodynamic laws governing these fluctuating macroscopic fields, we formulate stochastic thermodynamics at the hydrodynamic scale. We consider a system driven out of equilibrium by a time-dependent control protocol $\lambda_t$, which explicitly modulates the conservative confining potential $\Phi(\mathbf{r}, \lambda_t)$ \cite{sekimoto2010stochastic,jarzynski2012equalities,crooks1998nonequilibrium,seifert2012stochastic}.
Along a specific evolutionary trajectory of the macroscopic fields, formally denoted as $\boldsymbol{\Psi}_{[0,\tau]} := \{\rho_t, \mathbf{u}_t, s_t\}_{t \in [0,\tau]}$, the stochastic work $w[\boldsymbol{\Psi}_{[0,\tau]}]$ done on the system by the external agent is defined as
\begin{equation} \label{eq:stochastic_work}
	w[\boldsymbol{\Psi}_{[0,\tau]}] = \int_0^\tau dt \int \mathrm{d}\mathbf{r} \, \rho(\mathbf{r}, t) \frac{\partial \Phi(\mathbf{r}, \lambda_t)}{\partial \lambda_t} \dot{\lambda}_t.
\end{equation}
This expression purely captures the mechanical energy injected into the system exclusively through the controlled modulation of the potential $\Phi$. It is crucial to note that if the system is simultaneously driven by a non-conservative force $\mathbf{F}_{\mathrm{nc}}$, according to the paradigm of stochastic energetics, the work done by this force is accounted for as part of the stochastic heat exchange rather than the stochastic work \cite{seifert2012stochastic}.

Following the exact First Law of Thermodynamics at the trajectory level, the trajectory-dependent heat $q[\boldsymbol{\Psi}_{[0,\tau]}]$ exchanged with the reservoir is strictly defined as the difference between the change in total energy and the stochastic work
\begin{equation} \label{eq:stochastic_heat_def}
	q[\boldsymbol{\Psi}_{[0,\tau]}] \equiv E[\boldsymbol{\Psi}_\tau; \lambda_\tau] - E[\boldsymbol{\Psi}_0; \lambda_0] - w[\boldsymbol{\Psi}_{[0,\tau]}].
\end{equation}
Utilizing the local thermodynamic identity $\mathrm{d}u = T \mathrm{d}s + (p/\rho^2)\mathrm{d}\rho$, invoking the continuity equation~\eqref{eq:continuity}, and integrating by parts under appropriate boundary conditions\footnote{Specifically, the surface terms emerging from the spatial integration by parts identically vanish in two standard physical scenarios: (i) for a system under a confining potential, the macroscopic fluid density and fluxes naturally vanish at spatial infinity (or at the physical boundaries); (ii) under periodic boundary conditions, the fluid fields are identical at opposite boundaries, ensuring that the net boundary terms sum to exactly zero.}, the accumulated stochastic heat reduces to
\begin{equation} \label{eq:stochastic_heat_explicit}
	q[\boldsymbol{\Psi}_{[0,\tau]}] = \int_0^\tau \mathrm{d}t \int \mathrm{d}\mathbf{r} \left[ -\xi \rho \mathbf{u}^2 + \mathbf{u} \cdot \boldsymbol{\chi} - \gamma k_B \frac{\rho}{m} (T - T_r) + \Gamma \right].
\end{equation}
These four terms precisely capture the macroscopic frictional dissipation, the stochastic power injected by momentum fluctuations \cite{sekimoto2010stochastic}, the deterministic thermal relaxation, and the intrinsic fluctuating thermal exchange \cite{fei2024temperature}, respectively.

Having established the trajectory-level energy balance, we turn to the characterization of nonequilibrium irreversibility. Accounting for the probabilistic fluctuations of the macroscopic fields, we first define the stochastic system entropy $S_{\mathrm{sys}}[\boldsymbol{\Psi}_t]$ and the total entropy production $S_{\mathrm{tot}}[\boldsymbol{\Psi}_{[0,\tau]}]$ as
\begin{gather}
	S_{\mathrm{sys}}[\boldsymbol{\Psi}_t] = S[\boldsymbol{\Psi}_t] - k_B \ln P[\boldsymbol{\Psi}_t ; t], \label{eq:sys_entropy} \\
	S_{\mathrm{tot}}[\boldsymbol{\Psi}_{[0,\tau]}] = S_{\mathrm{sys}}[\boldsymbol{\Psi}_\tau] - S_{\mathrm{sys}}[\boldsymbol{\Psi}_0] - \frac{q[\boldsymbol{\Psi}_{[0,\tau]}]}{T_r}, \label{eq:tot_entropy}
\end{gather}
where $S[\boldsymbol{\Psi}_t] = \int \mathrm{d}\mathbf{r} \, \rho s$ represents the macroscopic Boltzmann entropy evaluated under the local equilibrium assumption, and $P[\boldsymbol{\Psi}_t ; t]$ explicitly denotes the instantaneous probability density of the hydrodynamic fields at time $t$. The term $-k_B \ln P[\boldsymbol{\Psi}_t ; t]$ constitutes the stochastic entropy distribution of the fluctuating fields, while $-q[\boldsymbol{\Psi}_{[0,\tau]}] / T_r$ quantifies the entropy change of the thermal reservoir.
While the energy $E[\boldsymbol{\Psi}_t; \lambda_t]$ is a state functional determined solely by the specific field configuration, the stochastic system entropy $S_{\mathrm{sys}}[\boldsymbol{\Psi}_t]$ is inherently statistical due to its explicit dependence on the instantaneous probability density $P[\boldsymbol{\Psi}_t ; t]$. Taking the ensemble average over all possible field configurations yields
\begin{equation}
	\langle S_{\mathrm{sys}}[\boldsymbol{\Psi}_t] \rangle = \langle S[\boldsymbol{\Psi}_t] \rangle - k_B \int \mathcal{D}\boldsymbol{\Psi} \, P[\boldsymbol{\Psi}; t] \ln P[\boldsymbol{\Psi}; t].
\end{equation}
From an information-theoretic perspective, this Shannon entropy contribution quantifies the macroscopic fluctuation uncertainty \cite{jaynes1957information,rosenkrantz2012jaynes}. As detailed in Appendix~\ref{app:Htheorem}, it is mathematically indispensable for proving the macroscopic H-theorem, i.e., the strict non-negativity of the average total entropy production, $\frac{d}{d\tau}\langle S_{\mathrm{tot}} \rangle \ge 0$. We would like to emphasize that, compared to overdamped models~\cite{leonard2013stochastic}, our approach extends field-level stochastic thermodynamics to the full hydrodynamic scale by retaining the momentum and entropy fields. This explicitly captures inertial dynamics and thermal relaxation intrinsically absent in the overdamped limit. 

We subsequently establish the fluctuation theorems for the macroscopic continuous fields. The conditional probability functional $\mathbb{P}[\boldsymbol{\Psi}_{[0,\tau]} | \boldsymbol{\Psi}_0; \lambda_{[0,\tau]}]$ of observing a specific forward trajectory is governed by both a rigid kinematic constraint and a stochastic thermodynamic weight, expressed as $\mathbb{P} \propto \delta(\partial_t \rho + \nabla \cdot (\rho \mathbf{u})) e^{-\mathcal{S}}$. Here, the Dirac $\delta$-functional strictly enforces deterministic macroscopic mass conservation, restricting the system to accessible density evolutions. Besides, the Onsager-Machlup functional $\mathcal{S}$ measures the likelihood of the fluid's momentum and thermal fluctuations. As derived in Appendix~\ref{app:action_derivation_from_FFPE}, this action over the time interval $[0,\tau]$ is explicitly given by
\begin{equation}
	\begin{aligned}
		\mathcal{S} &= \frac{1}{4 k_B T_r} \int_0^\tau dt \int \mathrm{d}\mathbf{r} \left[ \frac{\rho}{\xi} \left( \frac{\partial \mathbf{u}}{\partial t} + (\mathbf{u} \cdot \nabla)\mathbf{u} + \frac{1}{\rho}\nabla p + \nabla \Phi + \frac{\xi}{\rho} \frac{\delta F}{\delta \mathbf{u}} \right)^2 \right. \\
		&\quad + \left. \frac{\rho m T}{\gamma k_B} \left( \frac{\partial s}{\partial t} + \mathbf{u} \cdot \nabla s + \frac{\gamma k_B}{\rho m T} \frac{\delta F}{\delta s} \right)^2 \right].
	\end{aligned}
\end{equation}

To quantify the thermodynamic irreversibility, we compare the forward dynamics with a corresponding reverse process, which is explicitly designated by the superscript $R$. This reverse process is generated by executing the backward control protocol $\lambda^R_t \equiv \lambda_{\tau-t}$. Concurrently, the system evolves along the strictly time-reversed trajectory $\boldsymbol{\Psi}^R_t \equiv \boldsymbol{\Psi}^\dagger_{\tau-t}$, where the dagger notation ($\dagger$) denotes the parity transformation $\boldsymbol{\Psi}^\dagger = (\rho, -\mathbf{u}, s)^T$ necessary to account for the odd time-reversal parity of the macroscopic velocity field.
Evaluating the trajectory probability ratio under this time-reversal transformation, the macroscopic irreversibility is strictly dictated by the difference between the forward and reverse Onsager-Machlup functional, $\Delta \mathcal{S} = \mathcal{S} - \mathcal{S}^R$. Therefore, the detailed fluctuation theorem emerges (see Appendix~\ref{app:DTF} for the derivation)
\begin{equation} \label{eq:DFT}
	\ln \frac{\mathbb{P}[\boldsymbol{\Psi}_{[0,\tau]} | \boldsymbol{\Psi}_0; \lambda_{[0,\tau]}]}{\mathbb{P}^R[\boldsymbol{\Psi}^R_{[0,\tau]} | \boldsymbol{\Psi}^R_0; \lambda^R_{[0,\tau]}]} = \ln \frac{P_\tau[\boldsymbol{\Psi}_\tau]}{P_0[\boldsymbol{\Psi}_0]} + \frac{S_{\mathrm{tot}}[\boldsymbol{\Psi}_{[0,\tau]}]}{k_B}.
\end{equation}

By integrating over all possible trajectories and imposing a normalized probability distribution $P'_0[\boldsymbol{\Psi}^R_0]$ at the initiation of the reverse process, we obtain the generalized integral fluctuation theorem
\begin{equation} \label{eq:generalized_IFT}
	\left\langle \frac{P'_0[\boldsymbol{\Psi}^R_0]}{P_\tau[\boldsymbol{\Psi}_\tau]} e^{-S_{\mathrm{tot}}[\boldsymbol{\Psi}_{[0,\tau]}]/k_B} \right\rangle = 1.
\end{equation}

Choosing the specific reverse initial distribution $P'_0[\boldsymbol{\Psi}^R_0] = P_\tau[\boldsymbol{\Psi}_\tau]$ (i.e., initiating the reverse protocol directly from the true final distribution of the forward process) immediately recovers the standard integral fluctuation theorem for the total entropy production, $\langle e^{-S_{\mathrm{tot}}/k_B} \rangle = 1$ \cite{seifert2012stochastic}. From this exact non-equilibrium equality, the macroscopic Second Law of Thermodynamics, $\langle S_{\mathrm{tot}} \rangle \ge 0$, follows trivially via Jensen's inequality.

When both the forward and reverse processes are initiated from the quasi-equilibrium solutions of the functional Fokker-Planck equation (i.e., $P_0[\boldsymbol{\Psi}_0] = P_s[\boldsymbol{\Psi}_0; \lambda_0]$ and $P'_0[\boldsymbol{\Psi}^R_0] = P_s[\boldsymbol{\Psi}_\tau; \lambda_\tau]$) [Eq.~\eqref{eq:equilibrium_P}], the integral fluctuation theorem naturally yields the generalized Jarzynski equality and the corresponding principle of maximum work
\begin{equation} \label{eq:jarzynski}
	\langle e^{-\beta_r w[\boldsymbol{\Psi}_{[0,\tau]}]} \rangle = \frac{\mathcal{Z}(\lambda_\tau)}{\mathcal{Z}(\lambda_0)} = e^{-\beta_r \Delta \mathcal{F}}, \quad \langle w[\boldsymbol{\Psi}_{[0,\tau]}] \rangle \ge \Delta \mathcal{F}.
\end{equation}
Here, $\mathcal{Z}(\lambda)$ emerges as the functional normalization constant of the quasi-equilibrium distribution $P_s[\boldsymbol{\Psi}; \lambda]$, and $\Delta \mathcal{F} \equiv \mathcal{F}(\lambda_\tau) - \mathcal{F}(\lambda_0)$ represents the macroscopic quasi-equilibrium free energy difference between the final and initial stationary states. Crucially, evaluating statistical averages over this specific stationary ensemble demonstrates that $\mathcal{Z}(\lambda)$ acts strictly as the canonical partition function of the macroscopic fluid system, fully recovering its corresponding thermodynamic relations (for detailed results, see Ref.~\cite{fei2023quantum}).

\section{Applications to Quantum Fluids: Circuit Currents and BEC Criticality} \label{sec:applications}

\subsection{Electric Current in a Circuit} \label{subsec:circuit_current}

To study the transport properties of electrons, we consider an ideal electron gas under periodic boundary conditions. Crucially, we assume a spatially uniform system and no external confining potential ($\Phi_{\mathrm{ext}} = 0$). The system is driven by a uniform, time-dependent alternating current (AC) electric field $\mathbf{E}(t)$, corresponding to a non-conservative specific force $\mathbf{F}_{\mathrm{nc}}(t) = -e\mathbf{E}(t)/m$. Consequently, the macroscopic momentum balance equation is explicitly given by
\begin{equation} \label{eq:momentum_extF}
	\rho \left[ \frac{\partial \mathbf{u}}{\partial t} + (\mathbf{u} \cdot \nabla)\mathbf{u} \right] = -\nabla p - \rho \nabla \Phi_{\mathrm{int}} - \rho \frac{e}{m}\mathbf{E}(t) - \xi \rho \mathbf{u} + \boldsymbol{\chi}.
\end{equation}
Substituting the uniform electrical driving force $\mathbf{F}_{\mathrm{nc}}(t)$ into the functional Fokker-Planck equation~(\ref{eq:driven_FPE}), we obtain a solution for Eq.~(\ref{eq:driven_FPE}) given by
\begin{equation} \label{eq:transient_P}
	P[\boldsymbol{\Psi}; t] = \frac{1}{\widetilde{\mathcal{Z}}(t)} \exp\left\{ \frac{S[\boldsymbol{\Psi}]}{k_B} - \beta_r \int \mathrm{d}\mathbf{r} \, \rho \left[ \frac{1}{2} \big(\mathbf{u} - \langle \mathbf{u}(t) \rangle\big)^2 + \Phi_{\mathrm{int}} + u \right] \right\} \delta(N - N[\rho]).
\end{equation}
Here, $\widetilde{\mathcal{Z}}(t)$ is the time-dependent functional normalization constant, and $\langle \mathbf{u}(t) \rangle$ denotes the instantaneous macroscopic drift velocity, satisfying
\begin{equation} \label{eq:drift_electric}
	\frac{\mathrm{d}\langle \mathbf{u}(t) \rangle}{\mathrm{d}t} = -\xi \langle \mathbf{u}(t) \rangle - \frac{e}{m}\mathbf{E}(t).
\end{equation}
Transforming this macroscopic drift dynamics into the frequency domain naturally recovers the exact structure of the classical AC conductivity 
\begin{equation} \label{eq:ac_conductivity}
	\sigma(\omega) = \frac{n e^2}{m(\xi - i\omega)} = \frac{\sigma_0}{1 - i\omega\tau}.
\end{equation}
Here, $\sigma_0 = n e^2 / (m\xi)$ represents the static direct current (DC) electrical conductivity~\cite{ashcroft1976solid}, and $\tau \equiv \xi^{-1}$ defines the macroscopic relaxation time.

This derived macroscopic drift dynamics recovers the exact structure of the conventional Drude model \cite{kubo2012statistical}. However, the Drude model introduces $\xi$ strictly as a phenomenological constant, our functional formulation intrinsically redefines $\xi$ through the underlying microscopic statistical parameters of the fluctuating fields.
To explicitly evaluate this damping coefficient, we compute the isothermal derivative $\left(\partial \rho / \partial \mu\right)_T$ for a three-dimensional uniform ideal electron gas. Here, the electron energy reduces to the pure kinetic term $E = \mathbf{p}^2/2m$, where the zero-point is set at the bottom of the conduction band and $m$ denotes the effective mass. Under the local equilibrium assumption, the system obeys the standard Fermi-Dirac distribution $f(E) = \{e^{\beta(E - \mu)} + 1\}^{-1}$.

In the strong degeneracy regime ($k_B T \ll \mu$), the derivative $-\partial f / \partial E$ sharply peaks at the Fermi level $\mu$. Applying the standard Sommerfeld approximation \cite{ashcroft1976solid}, the isothermal density derivative simplifies directly to the density of states $\mathcal{N}$ at the Fermi surface

\begin{equation} \label{eq:density_derivative}
	\left( \frac{\partial \rho}{\partial \mu} \right)_T = m \int_0^\infty \mathcal{N}(E) \left( -\frac{\partial f}{\partial E} \right) \mathrm{d}E \approx m \mathcal{N}(\mu).
\end{equation}
Substituting this result back into our definition of the friction coefficient $\xi$, yields the explicit thermal scaling of the macroscopic damping

\begin{equation} \label{eq:friction_scaling}
	\xi \equiv \frac{\gamma k_B T}{\rho} \left(\frac{\partial \rho}{\partial \mu}\right)_T \approx \frac{\gamma k_B m}{\rho} \mathcal{N}(\mu) T.
\end{equation}
Consequently, the DC electrical resistivity $\sigma_0^{-1} = m\xi / (n e^2)$ evaluates precisely to
\begin{equation} \label{eq:resistivity_final}
	\sigma_0^{-1} = \frac{m \gamma k_B}{n^2 e^2} \mathcal{N}(\mu) T.
\end{equation}
Given the linear scaling of electric resistance $\sigma_0^{-1} \propto \mathcal{N}(\mu) T$, our phenomenological approach yields results consistent with the model established by Mott for phonon-induced electron scattering~\cite{mott1936electrical}. It is instructive to contrast our findings with the theoretical model in Ref.~\cite{chavanis2008nonlinear}, where the friction coefficient is introduced as a strict constant, lacking any intrinsic temperature dependence.

Our results can also be directly compared with the nonequilibrium thermodynamics of electrical conduction discussed by Bonança and Deffner~\cite{bonancca2022fluctuation}. In their AC-driven Drude model, the fluctuation theorem for irreversible entropy production is formulated based on macroscopic velocity fluctuations. In addition to capturing these velocity fluctuations, our framework further incorporates spatial dependence and temperature fluctuations into the stochastic entropy production.

\subsection{Stability and Critical Slowing Down near the Critical Temperature of BEC} \label{subsec:bec_criticality}

To investigate the macroscopic relaxation of a three-dimensional ($d=3$) Bose fluid as it approaches the critical temperature ($T_c$) from above, we assume the absence of any external confining potential ($\Phi_{\mathrm{ext}} = 0$). The fluid is thus subject solely to the internal interaction. For dilute Bose gases with isotropic short-range $s$-wave scattering, we explicitly evaluate this using the contact potential approximation $v(\mathbf{r} - \mathbf{r}') = gm \delta(\mathbf{r} - \mathbf{r}')$, where $g = 4\pi\hbar^2 a_s / m^3$ parameterizes the macroscopic interaction strength in terms of the $s$-wave scattering length $a_s$ \cite{dalfovo1999theory}. This interaction yields a local mean-field potential $\Phi_{\mathrm{int}} = g \rho$.
By substituting this local interaction into our macroscopic momentum equation \eqref{eq:momentum}, the corresponding driving force transforms exactly as $-\rho \nabla \Phi_{\mathrm{int}} = -\rho \nabla (g \rho) = -\nabla (\frac{1}{2}g\rho^2)$. Because this term combines additively with the kinetic pressure gradient $-\nabla p$, we can rigorously define the macroscopic interaction pressure
\begin{equation} \label{eq:p_int_final}
	p_{\mathrm{int}} = \frac{1}{2}g\rho^2.
\end{equation}
Consequently, the total driving force density in the fluid is given by the gradient of the total pressure, $-\nabla p_{\mathrm{tot}} = -\nabla (p + p_{\mathrm{int}})$. This effective pressure recovers the well-established thermodynamic equation of state \cite{huang1957quantum, pethick2002bose, lee1958low}. Finally, to prevent spinodal collapse, mechanical stability of the uniform fluid requires $\partial p_{\mathrm{tot}}/\partial \rho > 0$.

To simplify the analysis of the critical slowing down near $T_c$, we apply the overdamped limit. Specifically, when $\gamma$ is large, the left-hand side of Eq.~\eqref{eq:momentum} is a small quantity and can be neglected. Utilizing the local thermodynamic relation $\nabla p + \rho \nabla \Phi = \frac{\rho}{m} \nabla \mu$~\cite{marconi1999dynamic} and Eq.~\eqref{eq:momentum}, we obtain
\begin{equation} \label{eq:particle_flux}
	\rho \mathbf{u} = -\frac{\rho}{m\xi} \nabla \mu + \frac{\boldsymbol{\chi}}{\xi}.
\end{equation}
By substituting Eq.~\eqref{eq:particle_flux} into the continuity equation~\eqref{eq:continuity} and explicitly neglecting the stochastic noise term, we utilize the thermodynamic relations derived from the Bose-Einstein distribution $f = \left\{ \exp\left[ \beta \left( \frac{p^2}{2m} + g\rho - \mu \right) \right] - 1 \right\}^{-1}$, $\rho = \frac{m}{\lambda_T^3} \mathrm{Li}_{3/2}(z)$ and $\mu = k_B T \ln z + g\rho$, the evolution equation rigorously reduces to
\begin{equation} \label{eq:mu_evolution_physical}
	\frac{\partial \mu}{\partial t} = \left( \frac{k_B T \lambda_T^3}{m \mathrm{Li}_{1/2}(z)} + g \right) \frac{\partial}{\partial x} \left( \frac{\rho}{m\xi} \frac{\partial \mu}{\partial x} \right).
\end{equation}
where $\mathrm{Li}_n(z)$ denotes the polylogarithm function of order $n$, $z = \exp[(\mu - g\rho) / k_B T]$ is the local fugacity of the Bose liquid and  $\lambda_T = \sqrt{2\pi\hbar^2/(mk_B T)}$ is the thermal de Broglie wavelength. Here, the fluid is confined to a finite domain $x \in [0, L]$. To ensure particle conservation, the macroscopic particle flux must vanish at the boundaries, which naturally imposes the Neumann boundary conditions $\partial \mu / \partial x |_{x=0,L} = 0$.

Utilizing the effective friction coefficient $\xi = \gamma \mathrm{Li}_{1/2}(z)/\mathrm{Li}_{3/2}(z)$ and introducing the characteristic time $t_s = \gamma^{-1}$ alongside the thermal length $x_s = \sqrt{k_B T / (m\gamma^2)}$, we define the dimensionless variables $\bar{t} = t/t_s$, $\bar{x} = x/x_s$, $\bar{\mu} = \beta\mu$, $\bar{\rho} = \rho \lambda_T^3 / m$, and interaction strength $\bar{g} = m^2 g / (\lambda_T^3 k_B T)$. The deterministic governing equation then transforms into a dimensionless form
\begin{equation} \label{eq:dimensionless_evolution}
	\frac{\partial \bar{\mu}}{\partial \bar{t}} = \left( \frac{1}{\mathrm{Li}_{1/2}(z)} + \bar{g} \right) \frac{\partial}{\partial \bar{x}} \left[ \frac{\mathrm{Li}_{3/2}(z)^2}{\mathrm{Li}_{1/2}(z)} \frac{\partial \bar{\mu}}{\partial \bar{x}} \right].
\end{equation}

We investigate the critical slowing down phenomenon as the Bose fluid approaches the critical point $\bar{\mu}_c$, and to extract the corresponding relaxation time $\tau_R$. Conforming to the macroscopic no-flux Neumann boundary conditions, we introduce an initial spatial perturbation
\begin{equation} \label{eq:initial_perturbation}
	\bar{\mu}(\bar{x}, 0) = \bar{\mu}_c - \alpha - \alpha b \cos(\pi \bar{x}).
\end{equation}
Here, both $\alpha$ and $b$ are strictly small positive dimensionless parameters ($0 < \alpha \ll 1$, $0 < b \ll 1$). The parameter $\alpha$ explicitly quantifies the chemical potential difference from the critical point, while $b$ determines the relative amplitude of the initial spatial modulation. As the system evolves, it ultimately relaxes to a homogeneous final equilibrium state $\bar{\mu}_{\mathrm{eq}}$. Defining the local fluctuation as $\delta\bar{\mu}(\bar{x}, \bar{t}) \equiv \bar{\mu}(\bar{x}, \bar{t}) - \bar{\mu}_{\mathrm{eq}}$, we expand the equation~\eqref{eq:dimensionless_evolution} to the second order
\begin{equation} \label{eq:nonlinear_diffusion}
	\frac{\partial \delta\bar{\mu}}{\partial \bar{t}} = C_1(\bar{\mu}_{\mathrm{eq}}) \frac{\partial^2 \delta\bar{\mu}}{\partial \bar{x}^2} + C_2(\bar{\mu}_{\mathrm{eq}}) \left( \frac{\partial \delta\bar{\mu}}{\partial \bar{x}} \right)^2 + \mathcal{O}(\delta\bar{\mu}^3),
\end{equation}
where the effective linear diffusion coefficient $C_1$ and the second-order nonlinear convective coefficient $C_2$ are explicitly given by
\begin{align}
	C_1(\bar{\mu}_{\mathrm{eq}}) &= \left[ \mathrm{Li}_{1/2}(z_{\mathrm{eq}})^{-1} + \bar{g} \right] \frac{\mathrm{Li}_{3/2}(z_{\mathrm{eq}})^2}{\mathrm{Li}_{1/2}(z_{\mathrm{eq}})}, \\
	C_2(\bar{\mu}_{\mathrm{eq}}) &= \frac{\mathrm{Li}_{3/2}(z_{\mathrm{eq}})}{\mathrm{Li}_{1/2}(z_{\mathrm{eq}})^3} \left[ 2 \mathrm{Li}_{1/2}(z_{\mathrm{eq}})^2 - \mathrm{Li}_{3/2}(z_{\mathrm{eq}}) \mathrm{Li}_{-1/2}(z_{\mathrm{eq}}) \right].
\end{align}
As detailed in Appendix~\ref{app:asymptotic_scaling}, exact asymptotic analysis evaluated at the background equilibrium state near the quantum critical point ($\alpha \to 0$) yields $C_1(\bar{\mu}_{\mathrm{eq}}) \sim \mathcal{O}(\alpha)$ and $C_2(\bar{\mu}_{\mathrm{eq}}) \sim \mathcal{O}(1)$, dictating that the magnitude ratio of the second term to the first term in Eq.~\eqref{eq:nonlinear_diffusion} strictly scales as $\mathcal{O}(b)$. Therefore, restricting the perturbation amplitude to $b \ll 1$ safely truncates the nonlinear convection term and ensures $\bar{\mu}_{\mathrm{eq}} \approx \bar{\mu}_c - \alpha$ to leading order (as proved in Appendix~\ref{app:asymptotic_scaling}). Consequently, the full dynamics gracefully reduce to the purely linearized diffusion equation
\begin{equation} \label{eq:linearized_diffusion}
	\frac{\partial \delta\bar{\mu}}{\partial \bar{t}} \approx C_1(\bar{\mu}_{\mathrm{eq}}) \frac{\partial^2 \delta\bar{\mu}}{\partial \bar{x}^2}.
\end{equation}

To quantify the macroscopic relaxation time $\tau_R$, we utilize the non-equilibrium free energy functional $F(\bar{t})$ as our primary observable. This choice is physically dictated by the H-theorem\footnote{While the H-theorem in Section III was formulated in terms of the total entropy $S_{\mathrm{tot}}$, it is straightforward to show that in the absence of external work and stochastic noise, $\frac{\mathrm{d}}{\mathrm{d}\bar{t}} F \le 0$.}, which guarantees that $F(\bar{t})$ monotonically decays to its equilibrium minimum, thereby serving as a robust Lyapunov functional for the relaxation process. Under the linearized dynamics, the excess free energy $\delta F(\bar{t})$, defined as the difference between the instantaneous free energy and its equilibrium value, is found to decay exponentially (see Appendix~\ref{app:free_energy_relaxation} for the derivation)
\begin{equation} \label{eq:free_energy_variation}
	\delta F(\bar{t}) \propto \exp\left[ -2\pi^2 C_1(\bar{\mu}_{\mathrm{eq}}) \bar{t} \right].
\end{equation}
From this explicit temporal profile, we directly extract the inverse relaxation time as
\begin{equation} \label{eq:tau_def}
	\tau_R^{-1} = 2\pi^2 C_1(\bar{\mu}_{\mathrm{eq}}).
\end{equation}

\begin{figure}[b!]
	\centering
	\includegraphics[width=0.92\textwidth]{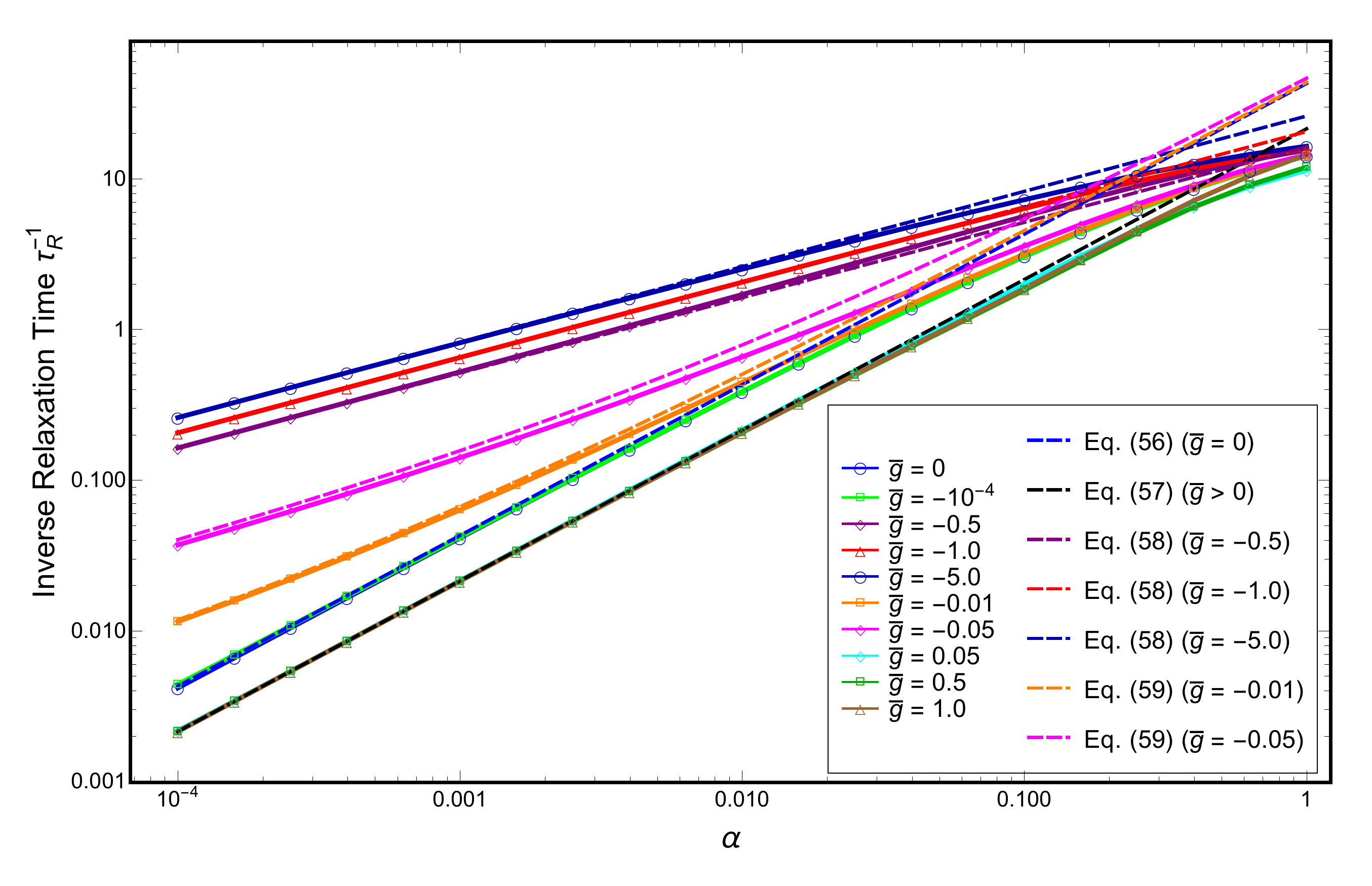}
	\caption{Inverse relaxation time $\tau_R^{-1}$ versus the chemical potential difference to criticality $\alpha$ across different interaction regimes. Solid lines with symbols denote numerical results extracted from the isothermal density evolution with an initial perturbation amplitude $b=0.01$. Dashed lines represent the corresponding analytical predictions: the ideal gas limit $\bar{g}=0$ [Eq.~\eqref{eq:tau_vanishing}], the repulsive regime $\bar{g}>0$ [Eq.~\eqref{eq:tau_repulsive}], the strong attractive regime (e.g., $\bar{g} \le -0.5$) following the $\mathcal{O}(\alpha^{1/2})$ spinodal scaling [Eq.~\eqref{eq:tau_finite}], and the weak attractive regime (e.g., $\bar{g} = -0.05, -0.01$) described by the composite asymptotic formula [Eq.~\eqref{eq:tau_composite}], which crosses over to the ideal gas scaling as $\alpha \to 0$. From top to bottom, all repulsive interactions ($\bar{g} > 0$) collapse onto the lowest branch, maintaining the linear $\mathcal{O}(\alpha)$ scaling of the ideal gas with half the prefactor.}
	\label{fig:universal_scaling}
\end{figure}

Our further analysis reveals that the dynamic scaling is fundamentally influenced by the interaction strength $\bar{g}$. In the ideal gas limit ($\bar{g} = 0$), driven purely by quantum fluctuations, the inverse relaxation time in Eq.~\eqref{eq:tau_def} exhibits a strict linear critical slowing down (see Appendix~\ref{app:scaling} for detailed derivations)
\begin{equation}\label{eq:tau_vanishing}
	\tau_R^{-1} \approx 2\pi \zeta(3/2)^2 \alpha.
\end{equation}
For repulsive interactions ($\bar{g} > 0$), the macroscopic flow is hindered by the interaction energy. This rigorously halves the relaxation prefactor in Eq.~\eqref{eq:tau_def} while preserving the linear $\mathcal{O}(\alpha)$ quantum scaling
\begin{equation} \label{eq:tau_repulsive}
	\tau_R^{-1} \approx \pi \zeta(3/2)^2 \alpha.
\end{equation}
Conversely, for attractive interactions ($\bar{g} < 0$), the system triggers a macroscopic spinodal collapse at a distinct critical threshold \cite{houbiers1996stability}. This instability point is characterized by a finite critical parameter $y_c = -\ln z_c$, which is uniquely determined by the spinodal condition $\mathrm{Li}_{1/2}(e^{-y_c}) = -1/\bar{g}$ (where the effective linear diffusion coefficient $C_1$ vanishes). Consequently, the dynamic scaling of Eq.~\eqref{eq:tau_def} is fundamentally shifted to a square-root dependence
\begin{equation} \label{eq:tau_finite}
	\tau_R^{-1} \approx 2\sqrt{2}\pi^2 \frac{\mathrm{Li}_{3/2}(e^{-y_c})^2 \sqrt{\mathrm{Li}_{-1/2}(e^{-y_c})}}{\mathrm{Li}_{1/2}(e^{-y_c})^{5/2}} \alpha^{1/2}.
\end{equation}
To smoothly bridge this finite attractive interaction regime and the ideal BEC limit ($\bar{g} \to 0^-$ and $y_c \to 0$), we employ matched asymptotic expansions~\cite{verhulst2005methods}. Additively composing the inner and outer limits of Eq.~\eqref{eq:tau_def} yields a unified composite scaling law valid across $\bar{g} < 0$ and $\bar{g} = 0$
\begin{equation} \label{eq:tau_composite}
	\tau_{R,\mathrm{com}}^{-1} \approx 2\sqrt{2}\pi^2 \frac{\mathrm{Li}_{3/2}(e^{-y_c})^2 \sqrt{\mathrm{Li}_{-1/2}(e^{-y_c})}}{\mathrm{Li}_{1/2}(e^{-y_c})^{5/2}} \alpha^{1/2} + 2\pi \zeta(3/2)^2 \alpha.
\end{equation}

Figure~\ref{fig:universal_scaling} compares our analytical predictions with numerical simulations. In the deep critical limit ($\alpha \ll 1$), the theoretical scaling laws match the numerical data. As $\alpha$ increases, neglected higher-order corrections emerge, causing deviations from these leading-order lines. Notably, the composite asymptotic matching [Eq.~\eqref{eq:tau_composite}] deviates at comparatively smaller values of $\alpha$. This earlier departure is expected, as structurally bridging two distinct asymptotic limits makes the composite expansion more sensitive to higher-order truncations.

\section{Conclusions and Outlook} \label{sec:conclusion}

By deriving macroscopic balance equations for fluctuating hydrodynamic fields from the phase-space NFPE, we have developed a phenomenological fluctuating hydrodynamics and a consistent stochastic thermodynamic framework, which encompasses the macroscopic thermodynamic laws and fluctuation theorems. We then applied this framework to investigate two macroscopic phenomena. For ideal Fermi gases, the dynamics reduces to a Drude model, revealing an electrical resistance proportional to the product of temperature and the density of states at the Fermi surface, consistent with both electron-phonon scattering theories~\cite{mott1936electrical} and experimental observations~\cite{ashcroft1976solid}. For overdamped bosonic fluids near the phase transition point, the local chemical potential obeys a linearized diffusion equation. The critical relaxation time scales linearly with the chemical potential difference to the critical point $\alpha$ in the non-interacting and repulsive regimes (differing by a factor of two), but follows a square-root scaling with respect to this distance in the attractive regime. Experimentally, our theory yields spatial density distributions that can potentially be verified using high-resolution optical imaging~\cite{bakr2009quantum}. Furthermore, the predicted critical scaling behaviors could potentially be tested in homogeneous ultracold bosonic gases confined by optical box potentials~\cite{gaunt2013bose}.

Finally, future work could extend this framework in several directions: incorporating non-diffusive mechanisms such as impurity scattering, and explicitly modeling the dynamics of the Bose-Einstein condensate below the critical temperature~\cite{zaremba1999dynamics}. Furthermore, fully capturing transport phenomena like viscosity and heat flux requires accounting for next-order deviations from the local equilibrium state, explicitly incorporating internal inter-particle collisions, which serve as the microscopic momentum exchange mechanisms driving macroscopic dissipation~\cite{chapman1970mathematical}.

\funding{This work was supported by the Quantum Science and Technology-National Science and Technology Major Project (Grant No.2024ZD0301000), the Science Challenge Project (No.TZ2025017), the National Natural Science Foundation of China (Grant No.12405046), and the Science Foundation of Zhejiang Sci-Tech University (Grants No.23062088-Y and No.23062181-Y).}

\appendixtitles{no} 
\appendixstart
\appendix
\section[\appendixname~\thesection]{Derivation of the Functional Fokker-Planck Equation}\label{app:FFP}
We first consider a fluid system confined by a conservative potential $\Phi$ in the absence of non-conservative external driving forces ($\mathbf{F}_{\mathrm{nc}} = \mathbf{0}$). To derive the functional Fokker-Planck equation, we define the probability functional $P[\rho, \mathbf{u}, s; t] = \langle \delta(\rho - \rho_t) \delta(\mathbf{u} - \mathbf{u}_t) \delta(s - s_t) \rangle$. Applying It\^o's lemma to the governing macroscopic Langevin equations \eqref{eq:continuity}, \eqref{eq:momentum}, and \eqref{eq:entropy}, and performing ensemble averaging yields its time evolution\cite{sekimoto2010stochastic}. By converting partial to functional derivatives via integration by parts and substituting the specific Langevin drifts, It\^o noise variances, alongside the free energy functional derivatives established in Eq.~\eqref{eq:F_derivatives}, the full equation algebraically expands as
\begin{equation} \label{eq:app_full_FP}
	\begin{aligned}
		\frac{\partial P}{\partial t} &= \int \mathrm{d}\mathbf{r} \nabla \cdot (\rho \mathbf{u}) \frac{\delta P}{\delta \rho} + \int \mathrm{d}\mathbf{r} \frac{1}{\rho} \left[ \rho (\mathbf{u} \cdot \nabla)\mathbf{u} + \nabla p + \rho \nabla \Phi + \xi \rho \mathbf{u} \right] \cdot \frac{\delta P}{\delta \mathbf{u}} \\
		&\quad + \int \mathrm{d}\mathbf{r} \frac{1}{\rho} \left[ \rho \mathbf{u} \cdot \nabla s + \frac{\gamma k_B \rho}{m} \frac{T - T_r}{T} \right] \frac{\delta P}{\delta s} + \int \mathrm{d}\mathbf{r} \frac{\xi k_B T_r}{\rho} \frac{\delta^2 P}{\delta \mathbf{u}^2} + \int \mathrm{d}\mathbf{r} \frac{\gamma k_B^2 T_r}{\rho m T} \frac{\delta^2 P}{\delta s^2} \\
		&= \int \mathrm{d}\mathbf{r} \left[ \nabla \cdot (\rho \mathbf{u}) \frac{\delta P}{\delta \rho} + \left( (\mathbf{u} \cdot \nabla)\mathbf{u} + \frac{\nabla p}{\rho} + \nabla \Phi \right) \cdot \frac{\delta P}{\delta \mathbf{u}} + (\mathbf{u} \cdot \nabla s) \frac{\delta P}{\delta s} \right] \\
		&\quad + \int \mathrm{d}\mathbf{r} \left( \frac{\xi}{\rho} \frac{\delta F}{\delta \mathbf{u}} \cdot \frac{\delta P}{\delta \mathbf{u}} + \frac{\xi k_B T_r}{\rho} \frac{\delta^2 P}{\delta \mathbf{u}^2} \right) + \int \mathrm{d}\mathbf{r} \left( \frac{\gamma k_B}{\rho m T} \frac{\delta F}{\delta s} \frac{\delta P}{\delta s} + \frac{\gamma k_B^2 T_r}{\rho m T} \frac{\delta^2 P}{\delta s^2} \right) \\
		&= \int \mathrm{d}\mathbf{r} \left[ - \frac{\delta F}{\delta \mathbf{u}} \cdot \nabla \frac{\delta P}{\delta \rho} + \frac{\delta P}{\delta \mathbf{u}} \cdot \nabla \frac{\delta F}{\delta \rho} - \frac{\delta F}{\delta s} \frac{\nabla s}{\rho} \cdot \frac{\delta P}{\delta \mathbf{u}} + \frac{\delta P}{\delta s} \frac{\nabla s}{\rho} \cdot \frac{\delta F}{\delta \mathbf{u}} \right] \\
		&\quad + \int \mathrm{d}\mathbf{r} \frac{\delta}{\delta \mathbf{u}} \cdot \left[ \frac{\xi P}{\rho} \left( \frac{\delta F}{\delta \mathbf{u}} + k_B T_r \frac{\delta \ln P}{\delta \mathbf{u}} \right) \right] + \int \mathrm{d}\mathbf{r} \frac{\delta}{\delta s} \left[ \frac{\gamma k_B P}{\rho m T} \left( \frac{\delta F}{\delta s} + k_B T_r \frac{\delta \ln P}{\delta s} \right) \right] \\
		&= \int \mathrm{d}\mathbf{r} \, \{F, P\} + \int \mathrm{d}\mathbf{r} \left( \frac{\delta}{\delta \mathbf{u}} \cdot \frac{\xi P}{\rho} \frac{\delta}{\delta \mathbf{u}} + \frac{\delta}{\delta s} \frac{\gamma k_B P}{\rho m T} \frac{\delta}{\delta s} \right) \left( F + k_B T_r \ln P \right).
	\end{aligned}
\end{equation}

Furthermore, consider a driven system in the absence of an external confining potential ($\Phi_{\mathrm{ext}} = 0$) but subjected to a uniform non-conservative external force $\mathbf{F}_{\mathrm{nc}}$. Incorporating the corresponding Langevin momentum dynamics given in Eq.~\eqref{eq:momentum_extF}, the functional Fokker-Planck equation expands as
\begin{equation} \label{eq:app_driven_FFPE_detailed}
	\begin{aligned}
		\frac{\partial P}{\partial t} &= \int \mathrm{d}\mathbf{r} \nabla \cdot (\rho \mathbf{u}) \frac{\delta P}{\delta \rho} + \int \mathrm{d}\mathbf{r} \frac{1}{\rho} \left[ \rho (\mathbf{u} \cdot \nabla)\mathbf{u} + \nabla p + \rho \nabla \Phi_{\mathrm{int}} - \rho \mathbf{F}_{\mathrm{nc}} + \xi \rho \mathbf{u} \right] \cdot \frac{\delta P}{\delta \mathbf{u}} \\
		&\quad + \int \mathrm{d}\mathbf{r} \frac{1}{\rho} \left[ \rho \mathbf{u} \cdot \nabla s + \frac{\gamma k_B \rho}{m} \frac{T - T_r}{T} \right] \frac{\delta P}{\delta s} + \int \mathrm{d}\mathbf{r} \frac{\xi k_B T_r}{\rho} \frac{\delta^2 P}{\delta \mathbf{u}^2} + \int \mathrm{d}\mathbf{r} \frac{\gamma k_B^2 T_r}{\rho m T} \frac{\delta^2 P}{\delta s^2} \\
		&= \int \mathrm{d}\mathbf{r} \left[ \nabla \cdot (\rho \mathbf{u}) \frac{\delta P}{\delta \rho} + \left( (\mathbf{u} \cdot \nabla)\mathbf{u} + \frac{\nabla p}{\rho} + \nabla \Phi_{\mathrm{int}} \right) \cdot \frac{\delta P}{\delta \mathbf{u}} + (\mathbf{u} \cdot \nabla s) \frac{\delta P}{\delta s} \right] - \int \mathrm{d}\mathbf{r} \mathbf{F}_{\mathrm{nc}} \cdot \frac{\delta P}{\delta \mathbf{u}} \\
		&\quad + \int \mathrm{d}\mathbf{r} \left( \frac{\xi}{\rho} \frac{\delta F}{\delta \mathbf{u}} \cdot \frac{\delta P}{\delta \mathbf{u}} + \frac{\xi k_B T_r}{\rho} \frac{\delta^2 P}{\delta \mathbf{u}^2} \right) + \int \mathrm{d}\mathbf{r} \left( \frac{\gamma k_B}{\rho m T} \frac{\delta F}{\delta s} \frac{\delta P}{\delta s} + \frac{\gamma k_B^2 T_r}{\rho m T} \frac{\delta^2 P}{\delta s^2} \right) \\
		&= \int \mathrm{d}\mathbf{r} \left[ - \frac{\delta F}{\delta \mathbf{u}} \cdot \nabla \frac{\delta P}{\delta \rho} + \frac{\delta P}{\delta \mathbf{u}} \cdot \nabla \frac{\delta F}{\delta \rho} - \frac{\delta F}{\delta s} \frac{\nabla s}{\rho} \cdot \frac{\delta P}{\delta \mathbf{u}} + \frac{\delta P}{\delta s} \frac{\nabla s}{\rho} \cdot \frac{\delta F}{\delta \mathbf{u}} \right] - \int \mathrm{d}\mathbf{r} \mathbf{F}_{\mathrm{nc}} \cdot \frac{\delta P}{\delta \mathbf{u}} \\
		&\quad + \int \mathrm{d}\mathbf{r} \frac{\delta}{\delta \mathbf{u}} \cdot \left[ \frac{\xi P}{\rho} \left( \frac{\delta F}{\delta \mathbf{u}} + k_B T_r \frac{\delta \ln P}{\delta \mathbf{u}} \right) \right] + \int \mathrm{d}\mathbf{r} \frac{\delta}{\delta s} \left[ \frac{\gamma k_B P}{\rho m T} \left( \frac{\delta F}{\delta s} + k_B T_r \frac{\delta \ln P}{\delta s} \right) \right] \\
		&= \int \mathrm{d}\mathbf{r} \, \{F, P\} - \int \mathrm{d}\mathbf{r} \mathbf{F}_{\mathrm{nc}} \cdot \frac{\delta P}{\delta \mathbf{u}} + \int \mathrm{d}\mathbf{r} \left( \frac{\delta}{\delta \mathbf{u}} \cdot \frac{\xi P}{\rho} \frac{\delta}{\delta \mathbf{u}} + \frac{\delta}{\delta s} \frac{\gamma k_B P}{\rho m T} \frac{\delta}{\delta s} \right) \left( F + k_B T_r \ln P \right).
	\end{aligned}
\end{equation}

\section[\appendixname~\thesection]{Proof Of The Macroscopic $H$-Theorem}\label{app:Htheorem}
\setcounter{equation}{0}
\renewcommand{\theequation}{B\arabic{equation}}
We first establish the fundamental relation between the total entropy production rate, the work power, and the stochastic free energy. According to the definitions of the stochastic thermodynamic quantities, the average total entropy production rate is the sum of the system and reservoir entropy changes. Utilizing the Clausius relation for the reservoir $\frac{d}{dt}\langle S_{res} \rangle = -\langle \dot{q} \rangle / T_r$, the first law of thermodynamics $\langle \dot{q} \rangle = \frac{d}{dt}\langle E \rangle - \langle \dot{w} \rangle$, and the definition of the stochastic free energy $\langle \mathcal{F} \rangle = \langle E \rangle - T_r \langle S_{sys} \rangle$, we systematically expand the total entropy production rate as
\begin{equation} \label{eq:H_start}
	\begin{aligned}
		\frac{d}{dt}\langle S_{tot}[\boldsymbol{\Psi}_{[0,t]}] \rangle &= \frac{d}{dt}\langle S_{sys} \rangle + \frac{d}{dt}\langle S_{res} \rangle \\
		&= \frac{d}{dt}\langle S_{sys} \rangle - \frac{\langle \dot{q} \rangle}{T_r} \\
		&= \frac{d}{dt}\langle S_{sys} \rangle - \frac{1}{T_r} \left( \frac{d}{dt}\langle E \rangle - \langle \dot{w} \rangle \right) \\
		&= \frac{1}{T_r} \left( \langle \dot{w}[\boldsymbol{\Psi}_{[0,t]}] \rangle - \frac{d}{dt}\langle \mathcal{F}[\boldsymbol{\Psi}_t; \lambda_t] \rangle \right).
	\end{aligned}
\end{equation}

Substituting the explicit form of the work power $\langle \dot{w} \rangle = \int \mathcal{D}\boldsymbol{\Psi} P[\boldsymbol{\Psi}] \frac{\partial F[\boldsymbol{\Psi}; \lambda_t]}{\partial t}$ and the definition of the stochastic free energy $\mathcal{F}[\boldsymbol{\Psi}_t; \lambda_t] = F[\boldsymbol{\Psi}_t; \lambda_t] + k_B T_r \ln P[\boldsymbol{\Psi}]$ into Eq.~\eqref{eq:H_start}, we continuously evaluate the evolution rate as
\begin{equation}
	\begin{aligned}
		\frac{d}{dt}\langle S_{tot}[\boldsymbol{\Psi}_{[0,t]}] \rangle &= \frac{1}{T_r} \Bigg( \int \mathcal{D}\boldsymbol{\Psi} P[\boldsymbol{\Psi}] \frac{\partial F[\boldsymbol{\Psi}; \lambda_t]}{\partial t} - \frac{d}{dt} \int \mathcal{D}\boldsymbol{\Psi} P[\boldsymbol{\Psi}] \big( F[\boldsymbol{\Psi}; \lambda_t] + k_B T_r \ln P[\boldsymbol{\Psi}] \big) \Bigg) \\
		&= \frac{1}{T_r} \Bigg( \int \mathcal{D}\boldsymbol{\Psi} P[\boldsymbol{\Psi}] \frac{\partial F}{\partial t} \\
		&\quad - \int \mathcal{D}\boldsymbol{\Psi} \Bigg[ \frac{\partial P[\boldsymbol{\Psi}]}{\partial t} (F + k_B T_r \ln P) + P[\boldsymbol{\Psi}] \frac{\partial F}{\partial t} + k_B T_r \frac{\partial P[\boldsymbol{\Psi}]}{\partial t} \Bigg] \Bigg) \\
		&= -\frac{1}{T_r} \int \mathcal{D}\boldsymbol{\Psi} \frac{\partial P[\boldsymbol{\Psi}]}{\partial t} \big( F[\boldsymbol{\Psi}; \lambda_t] + k_B T_r \ln P[\boldsymbol{\Psi}] + k_B T_r \big) \\
		&= -\frac{1}{T_r} \int \mathcal{D}\boldsymbol{\Psi} \frac{\partial P[\boldsymbol{\Psi}]}{\partial t} \mathcal{F}[\boldsymbol{\Psi}; \lambda_t].
	\end{aligned}
\end{equation}
Here, the last step utilizes the global normalization of the probability functional; since total probability is conserved to unity, its time derivative intrinsically vanishes, i.e., $\int \mathcal{D}\boldsymbol{\Psi} \frac{\partial P[\boldsymbol{\Psi}]}{\partial t} k_B T_r = 0$.

Substituting the Eq.~\eqref{eq:functional_FPE} into the above result, and performing functional integration by parts, we have
\begin{equation}\label{eq:Htheorem}
	\begin{aligned}
		\frac{d}{dt}\langle S_{tot}[\boldsymbol{\Psi}_{[0,t]}] \rangle &= -\frac{1}{T_r} \int \mathcal{D}\boldsymbol{\Psi} \Bigg\{ \int \mathrm{d}\mathbf{r} \, \{F, P\} \\
		&\quad + \int \mathrm{d}\mathbf{r} \left[ \frac{\delta}{\delta \mathbf{u}} \cdot \left( \frac{\xi P}{\rho} \frac{\delta \mathcal{F}}{\delta \mathbf{u}} \right) + \frac{\delta}{\delta s} \left( \frac{\gamma k_B P}{\rho m T} \frac{\delta \mathcal{F}}{\delta s} \right) \right] \Bigg\} \mathcal{F}[\boldsymbol{\Psi}; \lambda_t] \\
		&= -\frac{1}{T_r} \int \mathcal{D}\boldsymbol{\Psi} \int \mathrm{d}\mathbf{r} \, \{F, P\} \mathcal{F}[\boldsymbol{\Psi}; \lambda_t] \\
		&\quad + \frac{1}{T_r} \int \mathcal{D}\boldsymbol{\Psi} \int \mathrm{d}\mathbf{r} \left[ \left( \frac{\xi P}{\rho} \frac{\delta \mathcal{F}}{\delta \mathbf{u}} \right) \cdot \frac{\delta \mathcal{F}}{\delta \mathbf{u}} + \left( \frac{\gamma k_B P}{\rho m T} \frac{\delta \mathcal{F}}{\delta s} \right) \frac{\delta \mathcal{F}}{\delta s} \right] \\
		&= \frac{1}{T_r} \int \mathcal{D}\boldsymbol{\Psi} P[\boldsymbol{\Psi}] \int \mathrm{d}\mathbf{r} \frac{1}{\rho} \left[ \xi \left( \frac{\delta \mathcal{F}}{\delta \mathbf{u}} \right)^2 + \frac{\gamma k_B}{m T} \left( \frac{\delta \mathcal{F}}{\delta s} \right)^2 \right].
	\end{aligned}
\end{equation}

Because the first term exactly equals 0 (see Appendix ), and the second term is non-negative since $\rho > 0$, $\xi \ge 0$, and $\gamma \ge 0$, we strictly obtain the macroscopic $H$-theorem $\frac{d}{dt}\langle S_{tot}[\boldsymbol{\Psi}_{[0,t]}] \rangle \ge 0$.

\section[\appendixname~\thesection]{Reversible Term In Eq.~\eqref{eq:Htheorem}}
\label{app:reversible_entropy}
\setcounter{equation}{0}
\renewcommand{\theequation}{C\arabic{equation}}

We prove that the reversible streaming dynamics contribute identically zero macroscopic entropy production
\begin{equation}
	\frac{1}{T_r} \int \mathcal{D}\boldsymbol{\Psi} \int \mathrm{d}\mathbf{r} \, \{F, P\} \mathcal{F}[\boldsymbol{\Psi}; \lambda_t] = 0.
\end{equation}

Performing spatial integration by parts on the $\delta P / \delta \rho$ term, and applying the thermodynamic identity $\nabla u + \nabla(p/\rho) = T \nabla s + \nabla p / \rho$, the reversible drift integral evaluates directly as
\begin{equation}\label{eq:J_rev}
	\begin{aligned}
		\int \mathrm{d}\mathbf{r} \{F, P\} &= \int \mathrm{d}\mathbf{r} \left[ \frac{\delta F}{\delta \mathbf{u}} \cdot \left( \frac{\nabla s}{\rho} \frac{\delta P}{\delta s} - \nabla \frac{\delta P}{\delta \rho} \right) - \frac{\delta P}{\delta \mathbf{u}} \cdot \left( \frac{\nabla s}{\rho} \frac{\delta F}{\delta s} - \nabla \frac{\delta F}{\delta \rho} \right) \right] \\
		&= \int \mathrm{d}\mathbf{r} \Bigg[ \rho \mathbf{u} \cdot \left( \frac{\nabla s}{\rho} \frac{\delta P}{\delta s} - \nabla \frac{\delta P}{\delta \rho} \right) \\
		&\qquad\quad - \frac{\delta P}{\delta \mathbf{u}} \cdot \left( \frac{\nabla s}{\rho} (\rho T) - \nabla \left( \frac{1}{2}\mathbf{u}^2 + u + \frac{p}{\rho} + \Phi \right) \right) \Bigg] \\
		&= \int \mathrm{d}\mathbf{r} \Bigg[ \nabla \cdot (\rho \mathbf{u}) \frac{\delta P}{\delta \rho} + (\mathbf{u} \cdot \nabla s) \frac{\delta P}{\delta s} \\
		&\qquad\quad - \frac{\delta P}{\delta \mathbf{u}} \cdot \left( T \nabla s - (\mathbf{u} \cdot \nabla)\mathbf{u} - \left( T \nabla s + \frac{\nabla p}{\rho} \right) - \nabla \Phi \right) \Bigg] \\
		&= \int \mathrm{d}\mathbf{r} \left[ \nabla \cdot (\rho \mathbf{u}) \frac{\delta P}{\delta \rho} + \left( (\mathbf{u} \cdot \nabla)\mathbf{u} + \frac{\nabla p}{\rho} + \nabla \Phi \right) \cdot \frac{\delta P}{\delta \mathbf{u}} + (\mathbf{u} \cdot \nabla s) \frac{\delta P}{\delta s} \right],
	\end{aligned}
\end{equation}
Substituting the quasi-equilibrium free energy functional $\mathcal{F} = F + k_B T_r \ln P$ and Eq.~\eqref{eq:J_rev}, the total entropy production explicitly splits into $I_1$ and $I_2$ 
\begin{equation}
	\frac{1}{T_r} \int \mathcal{D}\boldsymbol{\Psi} \int \mathrm{d}\mathbf{r} \, \{F, P\} \mathcal{F}[\boldsymbol{\Psi}; \lambda_t] = \frac{1}{T_r} (I_1 + I_2),
\end{equation}
Treating the bare free energy $F[\boldsymbol{\Psi}]$ as a global functional, performing functional integration by parts over the phase space transfers the derivatives from $P$ to $F$ (with boundary terms vanishing)
\begin{equation}
	\begin{aligned}
		I_1 &= \int \mathcal{D}\boldsymbol{\Psi} \int \mathrm{d}\mathbf{r} \Bigg[ \nabla \cdot (\rho \mathbf{u}) \frac{\delta P}{\delta \rho} \\
		&\qquad\quad + \left( (\mathbf{u} \cdot \nabla)\mathbf{u} + \frac{\nabla p}{\rho} + \nabla \Phi \right) \cdot \frac{\delta P}{\delta \mathbf{u}} + (\mathbf{u} \cdot \nabla s) \frac{\delta P}{\delta s} \Bigg] F \\
		&= - \int \mathcal{D}\boldsymbol{\Psi} P \int \mathrm{d}\mathbf{r} \Bigg[ \nabla \cdot (\rho \mathbf{u}) \frac{\delta F}{\delta \rho} \\
		&\qquad\quad + \left( (\mathbf{u} \cdot \nabla)\mathbf{u} + \frac{\nabla p}{\rho} + \nabla \Phi \right) \cdot \frac{\delta F}{\delta \mathbf{u}} + (\mathbf{u} \cdot \nabla s) \frac{\delta F}{\delta s} \Bigg] \\
		&= \int \mathcal{D}\boldsymbol{\Psi} P \int \mathrm{d}\mathbf{r} \left[ \rho \mathbf{u} \cdot \left( \nabla u - T \nabla s - \frac{p}{\rho^2} \nabla \rho \right) \right],
	\end{aligned}
\end{equation}
Applying the local thermodynamic identity $\mathrm{d}u = T\mathrm{d}s + (p/\rho^2)\mathrm{d}\rho$, the spatial gradient expression $\nabla u - T \nabla s - \frac{p}{\rho^2} \nabla \rho \equiv 0$, yielding $I_1 = 0$.
Using the functional chain rule $\ln P \frac{\delta P}{\delta \Psi_i} = \frac{\delta}{\delta \Psi_i} ( P \ln P - P )$ and applying functional integration by parts, the global functional $(P \ln P - P)$ legally factors outside the spatial integral
\begin{equation}
	\begin{aligned}
		I_2 &= k_B T_r \int \mathcal{D}\boldsymbol{\Psi} \int \mathrm{d}\mathbf{r} \Bigg[ \nabla \cdot (\rho \mathbf{u}) \frac{\delta}{\delta \rho} \\
		&\qquad\quad + \left( (\mathbf{u} \cdot \nabla)\mathbf{u} + \frac{\nabla p}{\rho} + \nabla \Phi \right) \cdot \frac{\delta}{\delta \mathbf{u}} + (\mathbf{u} \cdot \nabla s) \frac{\delta}{\delta s} \Bigg] ( P \ln P - P ) \\
		&= 0.
	\end{aligned}
\end{equation}
Since $I_1 = 0$ and $I_2 = 0$, we have $I_1 + I_2 = 0$, confirming identically zero thermodynamic dissipation from the reversible streaming part.

\section[\appendixname~\thesection]{Derivation of the Onsager-Machlup Function from the Functional Fokker-Planck Equation}
\label{app:action_derivation_from_FFPE}
\setcounter{equation}{0}
\renewcommand{\theequation}{D\arabic{equation}}

Let the macroscopic field vector be $\boldsymbol{\Psi}(\mathbf{r}) = (\rho(\mathbf{r}), \mathbf{u}(\mathbf{r}), s(\mathbf{r}))^T$. The mobility operators associated with the dissipative terms are identified as $M_\rho = 0$, $M_{\mathbf{u}} = \xi/\rho$, and $M_s = \gamma k_B / (\rho m T)$.
Based on the underlying non-canonical Poisson bracket structure, the reversible drifts for the independent field variables are explicitly evaluated as
\begin{equation} \label{eq:app_evaluated_brackets}
	\{F, \rho\} = \nabla \cdot (\rho \mathbf{u}), \quad \{F, s\} = \mathbf{u} \cdot \nabla s, \quad \{F, \mathbf{u}\} = (\mathbf{u} \cdot \nabla)\mathbf{u} + \frac{1}{\rho}\nabla p + \nabla \Phi.
\end{equation}

By substituting the free energy functional $F$ and the probability density functional $P$ into the definition of the Poisson bracket, and applying integration by parts to the density derivative term ($-\int \frac{\delta F}{\delta \mathbf{u}} \cdot \nabla \frac{\delta P}{\delta \rho} \mathrm{d}\mathbf{r} = \int (\nabla \cdot \frac{\delta F}{\delta \mathbf{u}}) \frac{\delta P}{\delta \rho} \mathrm{d}\mathbf{r}$), the spatial integral is algebraically rearranged into
\begin{equation} \label{eq:app_bracket_expansion}
	\int \mathrm{d}\mathbf{r} \{F, P\} = \int \mathrm{d}\mathbf{r} \left[ \left(\nabla \cdot \frac{\delta F}{\delta \mathbf{u}}\right) \frac{\delta P}{\delta \rho} + \left(\frac{\delta F}{\delta \mathbf{u}} \cdot \frac{\nabla s}{\rho}\right) \frac{\delta P}{\delta s} + \left(\nabla \frac{\delta F}{\delta \rho} - \frac{\nabla s}{\rho} \frac{\delta F}{\delta s}\right) \cdot \frac{\delta P}{\delta \mathbf{u}} \right].
\end{equation}
The coefficients enclosed in the parentheses precisely correspond to the reversible Poisson drifts of the individual fields derived in Eq.~\eqref{eq:app_evaluated_brackets}. This direct substitution yields
\begin{equation} \label{eq:app_chain_rule_result}
	\int \mathrm{d}\mathbf{r} \{F, P\} = \int \mathrm{d}\mathbf{r} \sum_{i \in \{\rho, \mathbf{u}, s\}} \{F, \Psi_i\} \frac{\delta P}{\delta \Psi_i}.
\end{equation}

Furthermore, by utilizing the functional product rule, the integral of the probability flux divergence expands into two components
\begin{equation} \label{eq:app_product_rule}
	\int \mathrm{d}\mathbf{r} \sum_{i \in \{\rho, \mathbf{u}, s\}} \frac{\delta}{\delta \Psi_i} \Big( \{F, \Psi_i\} P \Big) = \int \mathrm{d}\mathbf{r} \sum_{i \in \{\rho, \mathbf{u}, s\}} \{F, \Psi_i\} \frac{\delta P}{\delta \Psi_i} + \int \mathrm{d}\mathbf{r} P \sum_{i \in \{\rho, \mathbf{u}, s\}} \frac{\delta}{\delta \Psi_i} \{F, \Psi_i\}.
\end{equation}
For the phase-space divergence appearing in the second term on the right-hand side, we express it algebraically. The mixed second-order functional derivatives algebraically cancel each other out, and the self-derivatives of the field-dependent differential operators evaluate to zero
\begin{equation} \label{eq:app_divergence_proof}
	\sum_{i \in \{\rho, \mathbf{u}, s\}} \frac{\delta}{\delta \Psi_i} \{F, \Psi_i\} = \left[ (-\nabla \cdot) \frac{\delta^2 F}{\delta \rho \delta \mathbf{u}} + (-\nabla) \frac{\delta^2 F}{\delta \mathbf{u} \delta \rho} \right] + \left[ \left(-\frac{\nabla s}{\rho} \cdot\right) \frac{\delta^2 F}{\delta s \delta \mathbf{u}} + \left(\frac{\nabla s}{\rho}\right) \frac{\delta^2 F}{\delta \mathbf{u} \delta s} \right] = 0.
\end{equation}
Since the phase-space divergence is identically zero, the second term in the product rule expansion Eq.~\eqref{eq:app_product_rule} vanishes, thereby rigorously establishing the following probability flux identity
\begin{equation} \label{eq:app_flux_identity}
	\int \mathrm{d}\mathbf{r} \{F, P\} = \int \mathrm{d}\mathbf{r} \sum_{i \in \{\rho, \mathbf{u}, s\}} \frac{\delta}{\delta \Psi_i} \Big( \{F, \Psi_i\} P \Big).
\end{equation}

Finally, combining this exact identity with the functional chain rule for the logarithmic dissipative term $\frac{\delta}{\delta \Psi_i} (k_B T_r \ln P) = \frac{k_B T_r}{P} \frac{\delta P}{\delta \Psi_i}$, the original functional Fokker-Planck equation is seamlessly and precisely rewritten into the generalized divergence form
\begin{equation} \label{eq:app_generalized_FFPE}
	\frac{\partial P}{\partial t} = \int \mathrm{d}\mathbf{r} \sum_{i \in \{\rho, \mathbf{u}, s\}} \frac{\delta}{\delta \Psi_i} \left\{ \{F, \Psi_i\} P + M_i \left[ \frac{\delta F}{\delta \Psi_i} P + k_B T_r \frac{\delta P}{\delta \Psi_i} \right] \right\}.
\end{equation}

Corresponding to the diffusion operator $k_B T_r M_i (\delta P / \delta \Psi_i)$, a Gaussian white noise field $\boldsymbol{\zeta}(\mathbf{r}, t) = (0, \boldsymbol{\zeta}_{\mathbf{u}}, \zeta_s)^T$ is defined in the functional space with the covariance matrix $\langle \zeta_i(\mathbf{r}, t) \zeta_j(\mathbf{r}', t') \rangle = 2 k_B T_r M_i \delta_{ij} \delta(\mathbf{r}-\mathbf{r}')\delta(t-t')$. The probability functional for the Gaussian noise is given by
\begin{equation} \label{eq:app_noise_prob}
	\mathbb{P}[\boldsymbol{\zeta}_{[0,\tau]}] \propto \exp \left( -\frac{1}{4 k_B T_r} \int_0^\tau dt \int \mathrm{d}\mathbf{r} \left[ \frac{\rho}{\xi} \boldsymbol{\zeta}_{\mathbf{u}}^2 + \frac{\rho m T}{\gamma k_B} \zeta_s^2 \right] \right).
\end{equation}

To perform the probability measure transformation from the noise fields to the physical fields, Dirac $\delta$-functionals are introduced to enforce the dynamical equations of motion
\begin{equation} \label{eq:app_delta_transformation}
	\mathbb{P}[\boldsymbol{\Psi}_{[0,\tau]} | \boldsymbol{\Psi}_0] = \int \mathcal{D}\boldsymbol{\zeta} \, \mathbb{P}[\boldsymbol{\zeta}_{[0,\tau]}] \prod_{t,\mathbf{r}} \delta \left( \frac{\partial \boldsymbol{\Psi}}{\partial t} + \{F, \boldsymbol{\Psi}\} + M \frac{\delta F}{\delta \boldsymbol{\Psi}} - \boldsymbol{\zeta} \right).
\end{equation}

Assuming the It\^o discretization, the functional Jacobian is constant. Integrating over $\mathcal{D}\boldsymbol{\zeta}$, the $\delta$-functional for the density field ($M_\rho=0, \zeta_\rho=0$) remains unintegrated, strictly preserving the deterministic continuity constraint $\delta(\partial_t \rho + \nabla \cdot (\rho \mathbf{u}))$. The dynamical residuals for the velocity and entropy fields are explicitly evaluated as
\begin{align}
	\boldsymbol{\zeta}_{\mathbf{u}} &= \frac{\partial \mathbf{u}}{\partial t} + (\mathbf{u} \cdot \nabla)\mathbf{u} + \frac{1}{\rho}\nabla p + \nabla \Phi + \frac{\xi}{\rho} \frac{\delta F}{\delta \mathbf{u}}, \label{eq:app_residual_u} \\
	\zeta_s &= \frac{\partial s}{\partial t} + \mathbf{u} \cdot \nabla s + \frac{\gamma k_B}{\rho m T} \frac{\delta F}{\delta s}. \label{eq:app_residual_s}
\end{align}

Substituting these residuals back into the exponential weight of Eq.~\eqref{eq:app_noise_prob}, the path probability simplifies to $\mathbb{P}[\boldsymbol{\Psi}_{[0,\tau]}] \propto \delta(\partial_t \rho + \nabla \cdot (\rho \mathbf{u})) \exp \big( -\mathcal{S} \big)$. The final Onsager-Machlup functional is rigorously derived as
\begin{equation} \label{eq:app_final_action}
	\begin{aligned}
		\mathcal{S} &= \frac{1}{4 k_B T_r} \int_0^\tau dt \int \mathrm{d}\mathbf{r} \left[ \frac{\rho}{\xi} \left( \frac{\partial \mathbf{u}}{\partial t} + (\mathbf{u} \cdot \nabla)\mathbf{u} + \frac{1}{\rho}\nabla p + \nabla \Phi + \frac{\xi}{\rho} \frac{\delta F}{\delta \mathbf{u}} \right)^2 \right. \\
		&\quad + \left. \frac{\rho m T}{\gamma k_B} \left( \frac{\partial s}{\partial t} + \mathbf{u} \cdot \nabla s + \frac{\gamma k_B}{\rho m T} \frac{\delta F}{\delta s} \right)^2 \right].
	\end{aligned}
\end{equation}

\section[\appendixname~\thesection]{Derivation of the Detailed Fluctuation Theorems}\label{app:DTF}
\setcounter{equation}{0}
\renewcommand{\theequation}{E\arabic{equation}}

To evaluate the path probability ratio under the time-reversal transformation $\mathcal{T}$, we introduce the mapped time variable $t' = \tau - t$. We first specify the time-reversal parities of the fluctuating macroscopic fields evaluated at time $t$: the density and entropy are even variables ($\rho^R(t) = \rho(t')$, $s^R(t) = s(t')$), while the fluid velocity is an odd variable ($\mathbf{u}^R(t) = -\mathbf{u}(t')$). 

Evaluating the forward Onsager-Machlup function $\mathcal{S}$ along the time-reversed trajectory $\boldsymbol{\Psi}^R$, the time derivatives transform correspondingly as $\partial_t \mathbf{u}^R(t) = \partial_{t'} \mathbf{u}(t')$ and $\partial_t s^R(t) = -\partial_{t'} s(t')$ due to the chain rule $\partial_t = -\partial_{t'}$. Furthermore, the local functional derivative $\delta F/\delta \mathbf{u} = \rho \mathbf{u}$ is time-reversal odd, whereas $\delta F/\delta s = \rho T$ remains even. Consequently, by substituting $t'$ back with a dummy integration variable $t$, the time-reversed action $\mathcal{S}^R$ explicitly takes the form
\begin{equation}
	\begin{aligned}
		\mathcal{S}^R &= \frac{1}{4 k_B T_r} \int_0^\tau dt \int d\mathbf{r} \left[ \frac{\rho}{\xi} \left( \frac{\partial \mathbf{u}}{\partial t} + (\mathbf{u} \cdot \nabla)\mathbf{u} + \frac{1}{\rho}\nabla p + \nabla \Phi - \frac{\xi}{\rho} \frac{\delta F}{\delta \mathbf{u}} \right)^2 \right. \\
		&\qquad\qquad\qquad\qquad\quad + \left. \frac{\rho T}{\gamma k_B} \left( -\frac{\partial s}{\partial t} - \mathbf{u} \cdot \nabla s + \frac{\gamma k_B}{\rho T} \frac{\delta F}{\delta s} \right)^2 \right].
	\end{aligned}
\end{equation}

The path probability ratio is determined by the action difference $\Delta \mathcal{S} = \mathcal{S} - \mathcal{S}^R$. Upon subtraction, the purely quadratic reversible and dissipative terms strictly cancel each other out. Only the cross terms between the time-reversal odd and even components survive
\begin{equation}
	\begin{aligned}
		-\Delta \mathcal{S} &= -\frac{1}{k_B T_r} \int_0^\tau dt \int d\mathbf{r} \left[ \left( \frac{\partial \mathbf{u}}{\partial t} + (\mathbf{u} \cdot \nabla)\mathbf{u} + \frac{1}{\rho}\nabla p + \nabla \Phi \right) \cdot \frac{\delta F}{\delta \mathbf{u}} \right. \\
		&\qquad\qquad\qquad\qquad\quad + \left. \left( \frac{\partial s}{\partial t} + \mathbf{u} \cdot \nabla s \right) \frac{\delta F}{\delta s} \right].
	\end{aligned}
\end{equation}

By adding the identically zero term related to the deterministic continuity equation $\left( \frac{\partial \rho}{\partial t} + \nabla \cdot (\rho \mathbf{u}) \right) \frac{\delta F}{\delta \rho} \equiv 0$, we can separate the integral into a time-evolution part and a convective part
\begin{equation}
		-\Delta \mathcal{S} = -\frac{1}{k_B T_r} \int_0^\tau dt \left[ \int d\mathbf{r} \left( \frac{\partial \rho}{\partial t} \frac{\delta F}{\delta \rho} \right.\right.  \left.\left. + \frac{\partial \mathbf{u}}{\partial t} \cdot \frac{\delta F}{\delta \mathbf{u}} + \frac{\partial s}{\partial t} \frac{\delta F}{\delta s} \right) - I_3 \right],
\end{equation}
where $I_3$ represents the free energy change driven strictly by the reversible dynamics
\begin{equation}
	\begin{aligned}
		I_3 &= \int \mathrm{d}\mathbf{r} \Bigg\{ [-\nabla \cdot (\rho \mathbf{u})] \left( \frac{1}{2}\mathbf{u}^2 + u + \frac{p}{\rho} + \Phi - T_r s \right) \\
		&\quad + \left[ -(\mathbf{u} \cdot \nabla)\mathbf{u} - \frac{\nabla p}{\rho} - \nabla \Phi \right] \cdot (\rho \mathbf{u})  + [-\mathbf{u} \cdot \nabla s] \rho(T - T_r) \Bigg\} \\
		&= \int \mathrm{d}\mathbf{r} \left[ \rho \mathbf{u} \cdot \left( \nabla u + \frac{\nabla p}{\rho} - \frac{p \nabla \rho}{\rho^2} \right) - \mathbf{u} \cdot \nabla p - \rho T \mathbf{u} \cdot \nabla s \right] \\
		&= \int \mathrm{d}\mathbf{r} \rho \mathbf{u} \cdot \left( \nabla u - \frac{p}{\rho^2}\nabla \rho - T \nabla s \right) \\
		&= 0.
	\end{aligned}
\end{equation}

The remaining terms in $-\Delta \mathcal{S}$ are exactly the total time derivative of $F$, minus the explicit time-dependence of the external potential $\Phi(\mathbf{r}, t)$
\begin{equation}
		-\Delta \mathcal{S} = -\frac{1}{k_B T_r} \int_0^\tau dt \left( \frac{dF}{dt} - \int d\mathbf{r} \rho \frac{\partial \Phi}{\partial t} \right) = \frac{w - \Delta F}{k_B T_r},
\end{equation}
where $w = \int_0^\tau dt \int d\mathbf{r} \rho \frac{\partial \Phi}{\partial t}$ is the stochastic work done on the system.

Substituting $\Delta F = \Delta E - T_r \Delta S$ and the first law of stochastic thermodynamics $\Delta E = w + q$ along the trajectory, the generalized local detailed balance (LDB) condition is obtained
\begin{equation}
	\ln \frac{\mathbb{P}[\boldsymbol{\Psi}_{[0,\tau]}|\boldsymbol{\Psi}_0]}{\mathbb{P}^R[\boldsymbol{\Psi}^\dagger_{[0,\tau]}|\boldsymbol{\Psi}^\dagger_\tau]} = -\Delta \mathcal{S} = \frac{\Delta S}{k_B} - \frac{q}{k_B T_r}.
\end{equation}

According to the definitions of the system stochastic entropy $S_{sys} = S - k_B \ln P$ and the reservoir entropy change $S_{res} = -q/T_r$, the total entropy production along the trajectory is $S_{tot} = \Delta S - k_B \ln(P_\tau/P_0) - q/T_r$. Substituting this into the LDB condition yields the detailed fluctuation theorem
\begin{equation}
	\ln \frac{\mathbb{P}[\boldsymbol{\Psi}_{[0,\tau]}|\boldsymbol{\Psi}_0]}{\mathbb{P}^R[\boldsymbol{\Psi}_{[0,\tau]}^\dagger|\boldsymbol{\Psi}_0^\dagger]} = \ln \frac{P_\tau[\boldsymbol{\Psi}_\tau]}{P_0[\boldsymbol{\Psi}_0]} + \frac{S_{tot}[\boldsymbol{\Psi}_{[0,\tau]}]}{k_B}.
\end{equation}

\section[\appendixname~\thesection]{Appendix: Asymptotic Scaling of Evolution Coefficients Near Criticality}\label{app:asymptotic_scaling}
\setcounter{equation}{0}
\renewcommand{\theequation}{F\arabic{equation}}

Near the critical point ($z \to 1$), we define the infinitesimal parameter $y = -\ln z \to 0^+$. The relevant polylogarithm functions asymptotically expand as
\begin{align}
	\mathrm{Li}_{3/2}(e^{-y}) &= \zeta(3/2) - 2\sqrt{\pi}y^{1/2} + \mathcal{O}(y), \\
	\mathrm{Li}_{1/2}(e^{-y}) &= \sqrt{\pi}y^{-1/2} + \zeta(1/2) + \mathcal{O}(y^{1/2}), \\
	\mathrm{Li}_{-1/2}(e^{-y}) &= \frac{\sqrt{\pi}}{2}y^{-3/2} + \zeta(-1/2) + \mathcal{O}(y^{-1/2}).
\end{align}

By definition, the fugacity is given by $z = \exp(\bar{\mu} - \bar{g}\bar{\rho})$. Substituting the background density $\bar{\rho} = \mathrm{Li}_{3/2}(z)$, the chemical potential expands as $\bar{\mu}(y) = -y + \bar{g}\mathrm{Li}_{3/2}(e^{-y}) \approx -y + \bar{g} [ \zeta(3/2) - 2\sqrt{\pi}y^{1/2} ]$. At the critical point ($y = 0$), the chemical potential is strictly $\bar{\mu}_c = \bar{g}\zeta(3/2)$. Defining the reference background state $\bar{\mu}_0 = \bar{\mu}_c - \alpha$, the chemical potential difference $\alpha$ relative to the critical point is given by
\begin{equation}
	\alpha = \bar{\mu}_c - \bar{\mu}_0 \approx y + 2\bar{g}\sqrt{\pi}y^{1/2}.
\end{equation}

To evaluate the linear diffusion coefficient $C_1$, we substitute the leading-order expansions and simplify
\begin{equation}
	\begin{aligned}
		C_1 &= \left[ \frac{1}{\mathrm{Li}_{1/2}(z)} + \bar{g} \right] \frac{\mathrm{Li}_{3/2}(z)^2}{\mathrm{Li}_{1/2}(z)} \\
		&\approx \left[ \frac{1}{\sqrt{\pi}y^{-1/2}} + \bar{g} \right] \frac{\zeta(3/2)^2}{\sqrt{\pi}y^{-1/2}} \\
		&= \left( \frac{y^{1/2}}{\sqrt{\pi}} + \bar{g} \right) \frac{\zeta(3/2)^2}{\sqrt{\pi}} y^{1/2} \\
		&= \frac{\zeta(3/2)^2}{\pi} \left( y + \bar{g}\sqrt{\pi}y^{1/2} \right).
	\end{aligned}
\end{equation}
Comparing this result directly with the mapping equation for $\alpha$, we strictly obtain $C_1 \propto \alpha$, confirming the critical slowing down behavior $C_1 \sim \mathcal{O}(\alpha)$.

For the nonlinear convective coefficient $C_2$, substituting the expansions reveals the explicit cancellation of singular divergences
\begin{equation}
	\begin{aligned}
		C_2 &= \frac{\mathrm{Li}_{3/2}(z)}{\mathrm{Li}_{1/2}(z)^3} \left[ 2 \mathrm{Li}_{1/2}(z)^2 - \mathrm{Li}_{3/2}(z) \mathrm{Li}_{-1/2}(z) \right] \\
		&\approx \frac{\zeta(3/2)}{(\sqrt{\pi}y^{-1/2})^3} \left[ 2(\sqrt{\pi}y^{-1/2})^2 - \zeta(3/2)\left(\frac{\sqrt{\pi}}{2}y^{-3/2}\right) \right] \\
		&= \left( \frac{\zeta(3/2)}{\pi^{3/2}} y^{3/2} \right) \left[ 2\pi y^{-1} - \frac{\sqrt{\pi}}{2}\zeta(3/2)y^{-3/2} \right] \\
		&= \frac{2\zeta(3/2)}{\sqrt{\pi}} y^{1/2} - \frac{\zeta(3/2)^2}{2\pi}.
	\end{aligned}
\end{equation}
In the limit $y \to 0^+$, the term proportional to $y^{1/2}$ vanishes. The strongest divergence ($y^{-3/2}$) perfectly cancels with the prefactor ($y^{3/2}$), leaving only the finite constant $C_2 \approx -\frac{\zeta(3/2)^2}{2\pi}$, indicating $C_2 \sim \mathcal{O}(1)$.

Finally, given the initial perturbation $\delta\bar{\mu} \sim \mathcal{O}(\alpha b)$, the corresponding spatial gradients scale as $\partial_{\bar{x}}\delta\bar{\mu}, \partial_{\bar{x}}^2\delta\bar{\mu} \sim \mathcal{O}(\alpha b)$. The relative magnitude of the nonlinear convection to the linear diffusion scales exactly as
\begin{equation}
	\frac{C_2 (\partial_{\bar{x}}\delta\bar{\mu})^2}{C_1 \partial_{\bar{x}}^2\delta\bar{\mu}} \sim \frac{\mathcal{O}(1) (\alpha b)^2}{\mathcal{O}(\alpha) (\alpha b)} = \mathcal{O}(b).
\end{equation}

By the macroscopic no-flux boundary conditions, the total dimensionless particle number is strictly conserved during the temporal evolution
\begin{equation}
	\int_0^1 \bar{\rho}(\bar{\mu}(\bar{x}, \bar{t})) \, \mathrm{d}\bar{x} = \text{const}.
\end{equation}

At the initial time $\bar{t}=0$, the chemical potential is prepared as $\bar{\mu}(\bar{x}, 0) = \bar{\mu}_0 - \alpha b \cos(\pi \bar{x})$ with the reference background state $\bar{\mu}_0 = \bar{\mu}_c - \alpha$. Expanding the equation of state $\bar{\rho}(\bar{\mu})$ up to the second order around $\bar{\mu}_0$ yields
\begin{equation}
	\begin{aligned}
		\bar{\rho}(\bar{\mu}(\bar{x}, 0)) &\approx \bar{\rho}(\bar{\mu}_0) + \bar{\rho}'(\bar{\mu}_0) \left[ - \alpha b \cos(\pi \bar{x}) \right] + \frac{1}{2}\bar{\rho}''(\bar{\mu}_0) \left[ - \alpha b \cos(\pi \bar{x}) \right]^2 \\
		&= \bar{\rho}(\bar{\mu}_0) - \alpha b \bar{\rho}'(\bar{\mu}_0) \cos(\pi \bar{x}) + \frac{1}{2} \alpha^2 b^2 \bar{\rho}''(\bar{\mu}_0) \cos^2(\pi \bar{x}).
	\end{aligned}
\end{equation}

Integrating over the spatial domain $\bar{x} \in [0, 1]$, the first-order cosine term identically vanishes, yielding the total initial particle number
\begin{equation}
	\begin{aligned}
		N(0) &= \int_0^1 \bar{\rho}(\bar{\mu}(\bar{x}, 0)) \, \mathrm{d}\bar{x} \\
		&\approx \bar{\rho}(\bar{\mu}_0) + \frac{1}{4} \alpha^2 b^2 \bar{\rho}''(\bar{\mu}_0).
	\end{aligned}
\end{equation}

As $\bar{t} \to \infty$, the system relaxes to a homogeneous final equilibrium state $\bar{\mu}_{\mathrm{eq}} = \bar{\mu}_0 + \Delta\bar{\mu}$, where $\Delta\bar{\mu}$ characterizes the macroscopic thermodynamic shift. Expanding the final density yields the ultimate particle number
\begin{equation}
	\begin{aligned}
		N(\infty) &= \int_0^1 \bar{\rho}(\bar{\mu}_{\mathrm{eq}}) \, \mathrm{d}\bar{x} \\
		&\approx \bar{\rho}(\bar{\mu}_0) + \bar{\rho}'(\bar{\mu}_0) \Delta\bar{\mu}.
	\end{aligned}
\end{equation}

Equating the initial and final particle numbers ($N(0) = N(\infty)$) exactly isolates the macroscopic shift
\begin{equation}
	\begin{aligned}
		\bar{\rho}'(\bar{\mu}_0) \Delta\bar{\mu} &= \frac{1}{4} \alpha^2 b^2 \bar{\rho}''(\bar{\mu}_0) \\
		\implies \Delta\bar{\mu} &= \frac{1}{4} \frac{\bar{\rho}''(\bar{\mu}_0)}{\bar{\rho}'(\bar{\mu}_0)} \alpha^2 b^2.
	\end{aligned}
\end{equation}

Near the quantum critical point, the density derivatives are governed by the singular scaling $\bar{\rho}''(\bar{\mu}_0)/\bar{\rho}'(\bar{\mu}_0) \sim \mathcal{O}(\alpha^{-1})$. Substituting this critical scaling strictly bounds the magnitude of the shift
\begin{equation}
	\Delta\bar{\mu} \sim \mathcal{O}(\alpha^{-1}) \cdot (\alpha b)^2 = \mathcal{O}(\alpha b^2).
\end{equation}

Given that the relative perturbation amplitude is strictly small ($b \ll 1$), this nonlinear shift constitutes a negligible higher-order correction ($\mathcal{O}(\alpha b^2) \ll \mathcal{O}(\alpha)$). Consequently, the final equilibrium state perfectly converges back to the unperturbed reference background
\begin{equation}
	\begin{aligned}
		\bar{\mu}_{\mathrm{eq}} &= \bar{\mu}_0 + \Delta\bar{\mu} \\
		&= \bar{\mu}_c - \alpha + \mathcal{O}(\alpha b^2) \\
		&\approx \bar{\mu}_c - \alpha.
	\end{aligned}
\end{equation}

\section[\appendixname~\thesection]{TEMPORAL RELAXATION OF THE FREE ENERGY FUNCTIONAL}\label{app:free_energy_relaxation}
\setcounter{equation}{0}
\renewcommand{\theequation}{G\arabic{equation}}

As shown in the main text, in the small-perturbation limit $b \ll 1$, nonlinear terms are negligible and the fluctuation $\delta\bar{\mu}(\bar{x}, \bar{t}) = \bar{\mu}(\bar{x}, \bar{t}) - \bar{\mu}_{\mathrm{eq}}$ obeys the linear diffusion equation
\begin{equation}
	\frac{\partial \delta\bar{\mu}}{\partial \bar{t}} = C_1(\bar{\mu}_{\mathrm{eq}}) \frac{\partial^2 \delta\bar{\mu}}{\partial \bar{x}^2}.
\end{equation}
Under the Neumann boundary conditions ($\partial_{\bar{x}}\delta\bar{\mu} = 0$ at $\bar{x}=0,1$), the general solution is a Fourier cosine series, $\delta\bar{\mu}(\bar{x}, \bar{t}) = \sum_{n=0}^{\infty} A_n \cos(n\pi\bar{x}) \exp[-\lambda_{\mu,n}\bar{t}]$, with decay rates $\lambda_{\mu,n} = n^2 \pi^2 C_1(\bar{\mu}_{\mathrm{eq}})$. The uniform $n=0$ mode vanishes ($A_0 = 0$) because $\delta\bar{\mu}$ is measured relative to the final equilibrium state $\bar{\mu}_{\mathrm{eq}}$ under total particle number conservation. Since $\lambda_{\mu,n}$ scales quadratically with $n$, higher-order modes ($n \ge 2$) decay much faster than the $n=1$ mode. Consequently, the long-time relaxation is dominated by the fundamental mode ($n=1$)
\begin{equation} \label{eq:mu_evolution}
	\delta\bar{\mu}(\bar{x}, \bar{t}) \propto \cos(\pi \bar{x}) \exp\left[ -\pi^2 C_1(\bar{\mu}_{\mathrm{eq}}) \bar{t} \right].
\end{equation}

To evaluate the macroscopic relaxation, we expand the free energy functional $F = \int_0^1 \bar{f}(\bar{\rho}) \, \mathrm{d}\bar{x}$ to second order around the homogeneous equilibrium state $\bar{\rho}_{\mathrm{eq}}$
\begin{equation}
	F(\bar{t}) = F_{\mathrm{eq}} + \int_0^1 \bar{\mu}_{\mathrm{eq}} \delta\bar{\rho} \, \mathrm{d}\bar{x} + \frac{1}{2} \int_0^1 \left. \frac{\partial \bar{\mu}}{\partial \bar{\rho}} \right|_{\mathrm{eq}} (\delta\bar{\rho})^2 \, \mathrm{d}\bar{x},
\end{equation}
where $\delta\bar{\rho}(\bar{x}, \bar{t}) = \bar{\rho}(\bar{x}, \bar{t}) - \bar{\rho}_{\mathrm{eq}}$. By conservation of total particle number, $\int_0^1 \delta\bar{\rho} \, \mathrm{d}\bar{x} = 0$, so the first-order term vanishes.

For the second-order term, using the linearized relation $\delta\bar{\rho} \approx \bar{\rho}'(\bar{\mu}_{\mathrm{eq}}) \delta\bar{\mu}$ and the thermodynamic identity $(\partial \bar{\mu}/\partial \bar{\rho})_{\mathrm{eq}} = 1/\bar{\rho}'(\bar{\mu}_{\mathrm{eq}})$, the variation of the free energy $\delta F(\bar{t}) = F(\bar{t}) - F_{\mathrm{eq}}$ becomes
\begin{equation}
	\begin{aligned}
		\delta F(\bar{t}) &= \frac{1}{2} \int_0^1 \frac{1}{\bar{\rho}'(\bar{\mu}_{\mathrm{eq}})} \left[ \bar{\rho}'(\bar{\mu}_{\mathrm{eq}}) \delta\bar{\mu} \right]^2 \, \mathrm{d}\bar{x} \\
		&= \frac{1}{2} \bar{\rho}'(\bar{\mu}_{\mathrm{eq}}) \int_0^1 \left[ \delta\bar{\mu}(\bar{x}, \bar{t}) \right]^2 \, \mathrm{d}\bar{x}.
	\end{aligned}
\end{equation}
Since the prefactor $\bar{\rho}'(\bar{\mu}_{\mathrm{eq}})/2$ is independent of time, the temporal evolution satisfies $\delta F(\bar{t}) \propto \int_0^1 (\delta\bar{\mu})^2 \, \mathrm{d}\bar{x}$. Substituting Eq.~\eqref{eq:mu_evolution} into this expression and integrating over $\bar{x}$ yields
\begin{equation}
	\delta F(\bar{t}) \propto \left( \exp\left[ -\pi^2 C_1(\bar{\mu}_{\mathrm{eq}}) \bar{t} \right] \right)^2 = \exp\left[ -2\pi^2 C_1(\bar{\mu}_{\mathrm{eq}}) \bar{t} \right].
\end{equation}
Thus, the relaxation rate of the free energy is twice the decay rate of the chemical potential fluctuation.

\section[\appendixname~\thesection]{DERIVATION OF THE CRITICAL RELAXATION SCALING}\label{app:scaling}
\setcounter{equation}{0}
\renewcommand{\theequation}{H\arabic{equation}}
We evaluate the asymptotic behavior of the collective diffusion coefficient $C_1(\bar{\mu}_{\mathrm{eq}})$ near the critical point by defining the target equilibrium state as $\bar{\mu}_{\mathrm{eq}} = \bar{\mu}_c - \alpha$ ($\alpha \ll 1$). The Robinson asymptotic expansions for the polylogarithm functions near $z \to 1$ (with $y = -\ln z \to 0^+$) are
\begin{equation}
	\mathrm{Li}_{3/2}(e^{-y}) \approx \zeta(3/2) - 2\sqrt{\pi y}, \quad \mathrm{Li}_{1/2}(e^{-y}) \approx \sqrt{\frac{\pi}{y}}.
\end{equation}

\vspace{0.3cm}
\noindent \textbf{1. Repulsive Interacting Fluid ($\bar{g} > 0$)}

The critical chemical potential is shifted by interactions to $\bar{\mu}_c = \bar{g} \zeta(3/2)$. Equating the target state to the self-consistent relation $\bar{\mu}_{\mathrm{eq}} = -y + \bar{g}\mathrm{Li}_{3/2}(e^{-y})$ yields
\begin{equation}
	\bar{g}\zeta(3/2) - \alpha \approx -y + \bar{g}\left[ \zeta(3/2) - 2\sqrt{\pi y} \right].
\end{equation}
Because $y \ll y^{1/2}$ near criticality, the linear kinetic term $-y$ is neglected, giving $y^{1/2} \approx \alpha / (2\bar{g}\sqrt{\pi})$. The inverse susceptibility thus scales as $1/\mathrm{Li}_{1/2}(z) \approx \sqrt{y/\pi} = \alpha / (2\pi\bar{g})$. Noting that $\bar{g} \gg 1/\mathrm{Li}_{1/2}(z)$, the collective diffusion coefficient simplifies to
\begin{equation}
	C_1(\bar{\mu}_{\mathrm{eq}}) \approx \bar{g} \frac{\mathrm{Li}_{3/2}(z)^2}{\mathrm{Li}_{1/2}(z)} \approx \bar{g} \left[ \frac{\zeta(3/2)^2}{2\pi \bar{g}} \alpha \right] = \frac{\zeta(3/2)^2}{2\pi} \alpha.
\end{equation}
Substituting this into the macroscopic relaxation relation $\tau_R^{-1} = 2\pi^2 C_1(\bar{\mu}_{\mathrm{eq}})$ directly yields $\tau_R^{-1} \approx \pi \zeta(3/2)^2 \alpha$.

\vspace{0.3cm}
\noindent \textbf{2. Ideal Bose Gas ($\bar{g} = 0$)}

In the absence of interactions, the critical chemical potential is exactly $\bar{\mu}_c = 0$, directly yielding $y = \alpha$. The collective diffusion coefficient incorporates the inverse susceptibility as a squared prefactor
\begin{equation}
	C_1(\bar{\mu}_{\mathrm{eq}}) = \left( \frac{1}{\mathrm{Li}_{1/2}(z)} \right) \frac{\mathrm{Li}_{3/2}(z)^2}{\mathrm{Li}_{1/2}(z)} = \frac{\mathrm{Li}_{3/2}(z)^2}{\mathrm{Li}_{1/2}(z)^2}.
\end{equation}
Substituting the asymptotic limits for $y = \alpha$ gives
\begin{equation}
	C_1(\bar{\mu}_{\mathrm{eq}}) \approx \zeta(3/2)^2 \left( \sqrt{\frac{\alpha}{\pi}} \right)^2 = \frac{\zeta(3/2)^2}{\pi} \alpha.
\end{equation}
Applying $\tau_R^{-1} = 2\pi^2 C_1(\bar{\mu}_{\mathrm{eq}})$ consequently yields $\tau_R^{-1} \approx 2\pi \zeta(3/2)^2 \alpha$.

\vspace{0.3cm}
\noindent \textbf{3. Attractive Interacting Fluid ($\bar{g} < 0$)}

For attractive interactions ($\bar{g} < 0$), mechanical spinodal collapse occurs when the macroscopic susceptibility diverges ($\partial_y \bar{\mu}_{\mathrm{eq}} = 0$). By differentiating the equation of state $\bar{\mu}_{\mathrm{eq}} = -y + \bar{g}\mathrm{Li}_{3/2}(e^{-y})$, the critical point $y_c$ satisfies
\begin{equation}
	\bar{\mu}'(y_c) = -1 + \bar{g}\left[ -\mathrm{Li}_{1/2}(e^{-y_c}) \right] = 0 \quad \implies \quad \bar{g} = -\frac{1}{\mathrm{Li}_{1/2}(e^{-y_c})}.
\end{equation}
Substituting $\bar{g}$ back yields the exact forms for the effective chemical potential and the collective diffusion coefficient
\begin{equation} \label{eq:mu_C1_def}
	\bar{\mu}(y) = -y - \frac{\mathrm{Li}_{3/2}(e^{-y})}{\mathrm{Li}_{1/2}(e^{-y_c})}, \quad C_1(y) = \frac{\mathrm{Li}_{3/2}(e^{-y})^2}{\mathrm{Li}_{1/2}(e^{-y})} \left[ \frac{1}{\mathrm{Li}_{1/2}(e^{-y})} - \frac{1}{\mathrm{Li}_{1/2}(e^{-y_c})} \right].
\end{equation}

\vspace{0.2cm}
\noindent \textbf{Outer Expansion (Finite Interaction, $y_c \sim \mathcal{O}(1)$):}

Defining the distance $\alpha = \bar{\mu}_c - \bar{\mu}_{\mathrm{eq}}$, we perform a second-order Taylor expansion of $\bar{\mu}_{\mathrm{eq}}(y)$ around $y_c$
\begin{equation}
	\bar{\mu}_{\mathrm{eq}}(y_{\mathrm{eq}}) \approx \bar{\mu}_c + \bar{\mu}'(y_c)(y_{\mathrm{eq}} - y_c) + \frac{1}{2}\bar{\mu}''(y_c)(y_{\mathrm{eq}} - y_c)^2.
\end{equation}
Noting the relations $\bar{\mu}_{\mathrm{eq}} - \bar{\mu}_c = -\alpha$ and $\bar{\mu}'(y_c) = 0$, and analytically evaluating the second derivative $\bar{\mu}''(y_c) = -\mathrm{Li}_{-1/2}(e^{-y_c})/\mathrm{Li}_{1/2}(e^{-y_c})$, we obtain
\begin{equation}
	-\alpha \approx \frac{1}{2} \left[ -\frac{\mathrm{Li}_{-1/2}(e^{-y_c})}{\mathrm{Li}_{1/2}(e^{-y_c})} \right] (y_{\mathrm{eq}} - y_c)^2 \quad \implies \quad y_{\mathrm{eq}} - y_c \approx \sqrt{ \frac{2 \mathrm{Li}_{1/2}(e^{-y_c})}{\mathrm{Li}_{-1/2}(e^{-y_c})} } \alpha^{1/2}.
\end{equation}
Since $C_1(y_c) = 0$, the leading-order expansion $C_1^{\mathrm{out}}(\alpha) \approx C_1'(y_c) (y_{\mathrm{eq}} - y_c)$ requires only the derivative of the bracketed term in Eq.~\eqref{eq:mu_C1_def}
\begin{equation}
	C_1'(y_c) = \frac{\mathrm{Li}_{3/2}(e^{-y_c})^2}{\mathrm{Li}_{1/2}(e^{-y_c})} \cdot \left[ \frac{\mathrm{Li}_{-1/2}(e^{-y_c})}{\mathrm{Li}_{1/2}(e^{-y_c})^2} - 0 \right] = \frac{\mathrm{Li}_{3/2}(e^{-y_c})^2 \mathrm{Li}_{-1/2}(e^{-y_c})}{\mathrm{Li}_{1/2}(e^{-y_c})^3}.
\end{equation}
Multiplying $C_1'(y_c)$ by $(y_{\mathrm{eq}} - y_c)$ explicitly yields the outer expansion
\begin{equation}
	C_1^{\mathrm{out}}(\alpha) \approx \left[ \sqrt{2} \frac{\mathrm{Li}_{3/2}(e^{-y_c})^2 \sqrt{\mathrm{Li}_{-1/2}(e^{-y_c})}}{\mathrm{Li}_{1/2}(e^{-y_c})^{5/2}} \right] \alpha^{1/2}.
\end{equation}

\vspace{0.2cm}
\noindent \textbf{Inner Expansion (Quantum Critical Limit, $y_c \to 0$):}

Applying the Robinson asymptotics $\mathrm{Li}_{1/2}(e^{-y}) \approx \sqrt{\pi/y}$ and $\mathrm{Li}_{3/2}(e^{-y}) \approx \zeta(3/2) - 2\sqrt{\pi y}$ to $\bar{\mu}(y)$ gives
\begin{equation}
	\bar{\mu}(y) \approx -y - \sqrt{\frac{y_c}{\pi}} \left[ \zeta(3/2) - 2\sqrt{\pi y} \right] = -y - \sqrt{\frac{y_c}{\pi}}\zeta(3/2) + 2\sqrt{y_c y}.
\end{equation}
Evaluating the exact difference $\bar{\mu}_{\mathrm{eq}} - \bar{\mu}_c = -\alpha$ cleanly cancels the $\zeta(3/2)$ constant term
\begin{equation}
	-\alpha \approx \left( -y_{\mathrm{eq}} + 2\sqrt{y_c y_{\mathrm{eq}}} \right) - \left( -y_c + 2\sqrt{y_c y_c} \right) = - (\sqrt{y_{\mathrm{eq}}} - \sqrt{y_c})^2.
\end{equation}
Taking the square root yields $\sqrt{y_{\mathrm{eq}}} = \sqrt{\alpha} + \sqrt{y_c}$. Applying the same Robinson asymptotics to $C_1(y)$ gives
\begin{equation}
	C_1(y_{\mathrm{eq}}) \approx \zeta(3/2)^2 \left[ \frac{y_{\mathrm{eq}}}{\pi} - \frac{\sqrt{y_{\mathrm{eq}} y_c}}{\pi} \right] = \frac{\zeta(3/2)^2}{\pi} \sqrt{y_{\mathrm{eq}}} (\sqrt{y_{\mathrm{eq}}} - \sqrt{y_c}).
\end{equation}
Substituting $\sqrt{y_{\mathrm{eq}}} - \sqrt{y_c} = \sqrt{\alpha}$ algebraically produces the inner expansion
\begin{equation}
	C_1^{\mathrm{in}}(\alpha) \approx \frac{\zeta(3/2)^2}{\pi} (\sqrt{\alpha} + \sqrt{y_c}) \sqrt{\alpha} = \frac{\zeta(3/2)^2}{\pi} (\alpha + \sqrt{\alpha y_c}).
\end{equation}

\vspace{0.2cm}
\noindent \textbf{Matched Asymptotic Unification:}

We extract the overlap limit ($y_c \to 0$) of the outer expansion prefactor by utilizing the singular asymptotic $\mathrm{Li}_{-1/2}(e^{-y_c}) \approx \frac{\sqrt{\pi}}{2} y_c^{-3/2}$
\begin{equation}
	C_1^{\mathrm{lap}}(\alpha) = \lim_{y_c \to 0} C_1^{\mathrm{out}}(\alpha) \approx \left[ \sqrt{2} \frac{\zeta(3/2)^2 \sqrt{\frac{\sqrt{\pi}}{2} y_c^{-3/2}}}{(\sqrt{\pi} y_c^{-1/2})^{5/2}} \right] \alpha^{1/2} = \frac{\zeta(3/2)^2}{\pi} \sqrt{\alpha y_c}.
\end{equation}
Applying additive composition $C_1^{\mathrm{com}} = C_1^{\mathrm{out}} + C_1^{\mathrm{in}} - C_1^{\mathrm{lap}}$ precisely cancels the cross term $\sqrt{\alpha y_c}$, preventing double-counting and yielding the globally valid unified scaling
\begin{equation}
	C_1^{\mathrm{com}}(\alpha) \approx \left[ \sqrt{2} \frac{\mathrm{Li}_{3/2}(e^{-y_c})^2 \sqrt{\mathrm{Li}_{-1/2}(e^{-y_c})}}{\mathrm{Li}_{1/2}(e^{-y_c})^{5/2}} \right] \alpha^{1/2} + \frac{\zeta(3/2)^2}{\pi} \alpha.
\end{equation}
Multiplying the unified collective diffusion coefficient by the fundamental mode factor $2\pi^2$ directly yields the composite inverse relaxation time presented in the main text
\begin{equation}
	\tau_{R,\mathrm{com}}^{-1} \approx 2\sqrt{2}\pi^2 \frac{\mathrm{Li}_{3/2}(e^{-y_c})^2 \sqrt{\mathrm{Li}_{-1/2}(e^{-y_c})}}{\mathrm{Li}_{1/2}(e^{-y_c})^{5/2}} \alpha^{1/2} + 2\pi \zeta(3/2)^2 \alpha.
\end{equation}

\begin{adjustwidth}{-\extralength}{0cm}

\reftitle{References}




\bibliographystyle{Definitions/mdpi}
\bibliography{myrefs}  

%


\PublishersNote{}
\end{adjustwidth}
\end{document}